\documentclass[preprint,preprint,12pt]{elsarticle}

\usepackage[top=2cm, bottom=2cm, right=2cm, left=2cm ]{geometry}
\usepackage{graphicx, subfigure}

\usepackage{epsfig}

\usepackage{amssymb}
\usepackage{amsthm}
\usepackage{amsmath}

\usepackage{lineno}
\usepackage{txfonts}
\usepackage{MnSymbol}

\usepackage{fancyhdr}
 
\journal{Applied Acoustics}

\begin{document}
\thispagestyle{empty}
This is the postprint of the paper: F. Auriemma, "Acoustic performance of micro-grooved elements", Applied Acoustics 122 (2017) 128-137, doi:10.1016/j.apacoust.2017.02.019.

\begin{frontmatter}
\lhead{Postprint of the paper: F. Auriemma, "Acoustic performance of micro-grooved elements", Applied Acoustics 122 (2017) 128-137, doi:10.1016/j.apacoust.2017.02.019.}



\title{}


\author{}

\address{}

\title{Acoustic Performance of Micro-Grooved Elements}

\author{Fabio Auriemma\corref{mycorrespondingauthor}}
\cortext[mycorrespondingauthor]{Corresponding author: Fabio Auriemma}
\ead{fabio.auriemma@ttu.ee}

\address{Department of Mechanical and Industrial Engineering, TT\"U - Tallinn University of Technology}

\address{Ehitajate tee 5, 19086, Tallinn, Estonia}

\begin{abstract}
The acoustic performance of a new type of fibre-less sound absorber, the Micro-Grooved Element (MGE), is studied in this paper. The transfer impedance and the absorption coefficient of micro-grooved and Micro-Perforated Elements (MPEs) are measured, modelled and compared. 

A MGE is a double layer element that involves inlet/outlet slots and facing micro-channels engraved onto the mating surface of the layers. The main advantage of these elements is that, by means of a simple technological process, thin micro-channels with depth of less then $100 \mu m$ can be easily engraved. This allows the MGEs exhibiting higher absorption coefficients compared to traditional MPEs with $300-700 \mu m$ diameter of perforations. Moreover, due to the presence of surfaces surrounding the micro-channels, the MGEs show reduced resistance when  exposed to high level of sound excitation. In this perspective, the performance of a MGE is more stable than the one of a MPE provided with the same porosity. 

A number of different MGEs with varied internal geometries has been tested at different excitation levels. The acoustics of the micro-channels is treated in details, the impedance end corrections are determined  and the non-linear effects are accounted for. The linear behaviour of MGEs is described by adapting the models for slit-shaped perforated elements  (SSEs). The \textit{quasi-} and \textit{non-} linear behaviours are expressed as a function of the Shear number and Strouhal number by curve fitting the experimental results.

\end{abstract}

\end{frontmatter}

\section{Introduction}\label{intro}

The recent trend in applied acoustics has seen gradual reduction of the use of fibrous materials and foams, especially in applications where the deterioration of these materials and the consequent pollution of the surrounding media is unwanted. Typical examples are HVAC systems \cite{Liu_Architectural_acoustics}, compressor silencers in supercharged engines \citep{Kabral_Compact_nonfibrous_silencer}, and turbine liners \cite{Bielak_Turbofan_Liner}.
Effective fibre-less solutions have been presented also in traditional automotive applications such as exhaust systems \cite{Allam_A_new_type_of_muff},  \cite{MGE_SETC_Application_to_small_eng_silencer}.

The strongest impulse to the diffusion of the fibre-less absorptive elements in the last two decades has been given by the advent of the micro-perforated elements MPEs \cite{Maa5_Basic_theory_of_acoustics}, \cite{Maa2_Theory_and_design_of_microperf}, \cite{Maa3_Microperf_at_high_sound}. In the MPEs the absorption occurs in apertures having diameters in the sub-millimeter range (typically in the range of $300-700 \mu m$ ) and perforation ratios of around $1\%$. In these conditions the Shear number within the apertures is of order unity, implying that the viscous boundary layer thickness is in the order of the perforation diameter. As a consequence, the dissipation of the acoustic energy is highly enhanced by means of the  viscous losses within the apertures. In substitution to the circular apertures, micro- slits with rectangular apertures can be utilized too \cite{Allard_book}. However, manufacturing costs of the MPEs can be an obstacle in engineering applications. In fact, a large number of the micro-apertures is required to provide adequate porosity, which makes the micro-perforation process technologically demanding and expensive.

Different designs and configurations of liners, based on the principle of the MPEs, have been proposed by researchers in the last fifteen years. The aim is typically the improvement of the acoustical performance of the MPEs and/or the reduction of the manufacturing costs. In some cases the solution proposed were able to pursue only one of the two goals, by penalizing the other one. 

Improvement of the absorption coefficient has been documented when the MPEs are utilized in combination with partitioned back cavities \cite{Liu_Enhancing_by_partitioning_cavity}, as well as in multiple leaf configurations \cite{Sagakami_Enhancing_by_multiple_leaf}, or backed by Helmholtz resonators \cite{Park_microperf_backed_by_Helmholtz}. In \cite{Quian_Ultra_micro_perforations}, the acoustic performance of typical MPEs has been improved by utilizing ultra-micro perforation with diameter less then $100 \mu m$ based on MEMS technology. In fact, as described by Maa in \cite{Maa2_Theory_and_design_of_microperf}, a straightforward and space-saving way to approach the broadband absorption is to reduce the perforation diameter and increase the perforation ratio adequately.  
In \cite{Randeberg_Perforated_absorbers}, the orifice design of circular perforation has been improved by using horn-shaped profiles in order to reduce the end effects of the apertures. In case of ultra-micro perforations and horn shaped micro-perforations, the technological issues involved in the production are even more critical than those required by traditional MPEs.

Manufacturing costs have been reduced by using different designs of the apertures, such as punched slit-shaped irregular apertures  \cite{Knipstein_Acoustimet}, micro-perforated insertion units (MIUs) \cite{Pfretzchner_Insertion_units}, and  double panel absorbers with small air gap \cite{Randeberg_Perforated_absorbers}, to cite a few examples. Among these solutions, the one proposed in \cite{Knipstein_Acoustimet} is currently utilized in engineering applications, despite the surface roughness of these elements and the reduced predictability of the performance.

In this paper, the acoustic mechanism of a new type of fibre-less acoustical absorber, the Micro-Grooved Element (MGE), is described. The acoustic performance is experimentally characterized in terms of transfer impedance and absorption coefficient. A complete model of the MGEs is developed by adapting the formulation for slit shaped apertures, by including the end corrections of the micro-channels and by relating the non-linear effects to the excitation amplitude and to the thickness of the oscillating viscous boundary layer. Presented for the first time in 2012 \cite{AuriemmaMGE1_A_novel_solution}, the initial design of the MGEs has been improved to optimize cost and performance. Since the MGEs are based on a number of micro-channels engraved on the surface of two slotted mating plates, the manufacturing process is technologically effective. Nevertheless, the depth of the micro-channels can be easily $< 100 \mu m$, as some of the ultra-micro perforated element presented in \cite{Quian_Ultra_micro_perforations}, while still preserving an adequate perforation ratio. Thus, the exhibited absorption coefficients can reach higher values than that of the traditional MPEs provided with $300-700 \mu m$ perforation diameter.  

As known from the literature, the non-linear effects exhibited at high level of sound excitation strongly affect the performance of the MPEs, \cite{Maa3_Microperf_at_high_sound}, \cite{Melling_Acoust_impedance_at_high_medium_pressure} \cite{Cummings_High_amplitude_transmission} \cite{Elnady_Semi_Empirical_liner_impedance_modeling}, \cite{Temiz_Nonlinear_acou_transfer}. In \cite{Park_microperf_launcher_fairings}, a design method for panel absorbers for high pressure environments is presented, based on the use of large hole apertures and in \cite{Chandrasekharan_Acoustic_impedance_of_MEMS_based_microperf} the MPEs have been designed with a high ratio length/diameter of the apertures in order to reduce the non-linear effects. However, both techniques penalize tremendously the performance at low excitation levels.

In case of MGEs the presence of surfaces surrounding the micro-channels and parallel to them, results in reduced acoustic resistance when these absorbers are exposed to high pressure levels. This implies that the acoustic performance of MGE, at varying the acoustic excitation, is more stable than the one of MPE provided with the same porosity. However, the techniques described in \cite{Park_microperf_backed_by_Helmholtz} and \cite{Chandrasekharan_Acoustic_impedance_of_MEMS_based_microperf}, can be still applied to MGEs.

Future studies are planned where the response modelling and the evolutionary optimization techniques are used to reduce the non-linear effects in a wider range of sound excitation levels \cite{Majak_genetic_algoritm}. 

\section{Transfer impedance}\label{Transf_Imp}
\subsection{Definitions}
When the acoustic waves pass through an intervening porous slab there is a net flow through it. If $\hat{u}_{a}$ and $\hat{u}_{b}$ are the particle velocities in front and at the back of the surface respectively, assuming the velocities $\hat{u}=\hat{u}_{a}=\hat{u}_{b}$ is a good approximation if the pore volume per unit slab area is substantially less than $1/4$ wavelength \cite{Pierce_book}. This is the case of the MGE, whose typical thickness is $\approx 1 mm$.
In this hypothesis one can define the \textit{transfer impedance} $Z_{tr}$ as follows:
\begin{equation}\label{def_Transfer_impedance}
\hat{p}_{a}-\hat{p}_{b}=Z_{tr}\hat{u}_{a}=Z_{tr}\hat{u}_{b}
\end{equation}

Here $\hat{p}_{a}$ and $\hat{p}_{b}$ represent the complex acoustic pressure amplitudes at the front and back sides of the slab. 
Dividing both sides by $\hat{u}$, and making use of the definition of \textit{specific acoustic impedance} on a surface $S_{0}$ as the ratio between the complex amplitudes of the acoustic pressure and the particle velocity measured on $S_0$, $ Z_{s_{0}}\left( \omega \right)=  \left( \hat{p}/\hat{u}_{in}\right)_{S_0}$, yields:

\begin{equation}\label{series_impedance}
Z_{a}=Z_{b}+Z_{tr}
\end{equation} 
In case of MGE the slab of material is constituted by an inlet layer ${In}$ , an outlet layer ${Out}$ and a number of micro-channels ${Ch}$ located on the mating surface of the two layers. In this case, the Eq. \eqref{series_impedance} becomes:

\begin{equation}\label{series_impedance_MGE}
Z_{tr}=Z_{In_{a}}-Z_{In_{b}}+Z_{m-ch_{a}}-Z_{m-ch_{b}}+Z_{Out_{a}}-Z_{Out_{b}}=Z_{tr-In}+Z_{tr-Ch}+Z_{tr-Out}
\end{equation} 
Where $Z_{tr-In}$, $Z_{tr-Ch}$ and $Z_{tr-Out}$ represent, respectively, the transfer impedances of the inlet layer, micro-channels and outlet layer considered separately.

For a perforated slab provided with several apertures, if $Z_{tr-s}$ is the transfer impedance of a single aperture, then the transfer impedance of the entire element is $Z_{tr} =Z_{tr-s}/ \sigma$, where $\sigma$ is the panel porosity. $Z_{tr}$ is a complex quantity, i.e. $Z_{tr}=R+ \mathrm{i}X$, where the real part $R$ (or resistive component) is referred to as transfer resistance and the imaginary part $X$ (or reactive component) is typically named transfer reactance. 
The transfer impedance and its components are usually normalized with respect to the air properties $\rho_0 c$ ($\rho_0$ is the air density $\approx 1.18 kg/m^3$ at standard temperature and pressure and $c$ is the speed of sound) and are here indicated with $z_{tr}$, $r$ and $\chi$.

\subsection{Transfer impedance measurement}
Samples studied in this work are listed in Table\ref{Tab_1}.
\begin{table}[ht!]
  \centering
  \begin{tabular}{c c c c c c}
    \hline
    \textit{Sample name} & $d$ $[\mu m]$ & $t$ $[\mu m]$ &  $C_{vc}$ & \textit{n.apertures} & \textit{porosity}   \\\hline 
MGE0  &  95  &  200 & $\approx 0.8$ & 8 & 1.52\\
MGE1  & 120 &  [250, 750, 1200, 1750] & $\approx 0.8$ & 8 & 1.19\\
MGE2  & 190 &  [250, 750, 1200, 1750] & $\approx$ 0.8 & 8 & 1.89\\
MGE3  & 275 &  [250, 750, 1200, 1750] & $\approx 0.8$ & 8 & 2.73\\
MGE4  & 255 &  [250, 750, 1200, 1750] & $\approx 0.8$ & 8 & 1.81\\
MGE5  & 455 &  [250, 750, 1200, 1750] & $\approx 0.8$ & 8 & 3.23\\\hline
MPE1  & 300 & 500  & $\approx 0.7$ & 300 & 1.53\\
MPE2  & 400 & 500  & $\approx 0.7$ & 120 & 1.09\\\hline
  \end{tabular}
  \caption{Data of the samples tested} \label{Tab_1}
\end{table}
Several methods for the determination of the transfer impedance of these elements can be found in literature, \cite{Allard_book}.
In this work, the 2-port data of the samples have been experimentally determined in order to carry out the associated transfer impedance, \cite{Lavrentjev_A_measurement_method}. The sound field propagating in the test-rig below the first cut-on frequency consists of plane waves. In the frequency domain, this condition is described as: 
\begin{equation}\label{aco_field_p}
\hat{p}(x,f)= \hat{p}_+(f)\exp(-\mathrm{i}k_+x)+\hat{p}_-(f)\exp(\mathrm{i}k_-x)    
\end{equation} 

\begin{equation}\label{aco_field_u}
\hat{u}(x,f)= \frac{1}{\rho_0 c} \left[  \hat{p}_+(f)\exp(-\mathrm{i}k_+x)-\hat{p}_-(f)\exp(\mathrm{i}k_-x) \right]      
\end{equation} 

\begin{figure} 
     \centering
        {\includegraphics[scale=0.7]{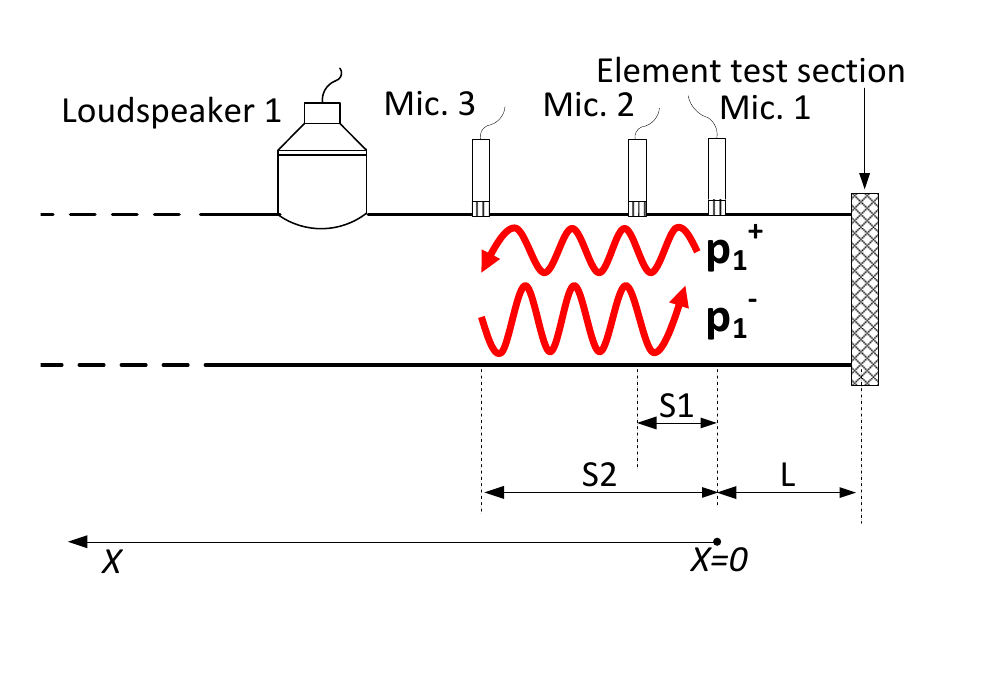}}

         \caption{Determination of the particle velocities}     
         \label{fig_TestRig_Vel}
\end{figure}

Here, conventionally, the x-directions are oriented outward the test object (see Figure \ref{fig_TestRig_Vel}), $\pm$ denotes the propagation in positive and negative x-directions, $f$ is the frequency and $k$ is the complex wavenumber. In absence of a mean flow, as the case in this study, $k_+=k_-=k_0$.  

By assuming the complex wavenumber known in the duct, the incident and reflected waves can be calculated, on both sides of the test object, with the classical wave decomposition technique \cite{Chung_Transfer_funct}. For example, at the position Mic.1 of the side $a$ (front) depicted in Fig. \ref{fig_TestRig_Vel}, one can write:
\begin{equation}\label{p_plus}
\hat{p}_{1+}= \frac{\hat{p}_1\exp(\mathrm{i}k_0x_2)-\hat{p}_2}   
                   {2\mathrm{i} \sin (k_0x_2)}       
\end{equation}
\begin{equation}\label{p_minus}
\hat{p}_{1-}= \frac{\hat{p}_2-\hat{p}_1\exp(-\mathrm{i}k_0s)}   
                   {2\mathrm{i} \sin (k_0x_2)}       
\end{equation}
where $\hat{p}_{1}$ and $\hat{p}_{2}$ are the amplitude components of the acoustic pressure in in the positions Mic.1 and Mic.2 (or Mic.1 and Mic.3).
In correspondence of the test section, $x=-L$, so that:

\begin{equation}\label{p_test-sect}
\hat{p}_{a}(-L,f)=  \hat{p}_{1+} \exp(\mathrm{i}k_0L) + \hat{p}_{1-} \exp(-\mathrm{i}k_0L) 
\end{equation}

\begin{equation}\label{u_test-sect}
\hat{u}_{a}(-L,f)= \frac{1}{\rho_0 c} \left[\hat{p}_{1+}\exp(\mathrm{i}k_0L)  - \hat{p}_{1-}\exp(-\mathrm{i}k_0L)\right] 
\end{equation}

At this point it is straightforward to compute the transfer matrix \textbf{T} of the test object, \cite{Munjal_book} which is defined as:
\begin{equation}\label{T_matrix}
\left(
     \begin{array}{c}
  \hat{p}_{b} \\
  \hat{q}_{b} \\
  \end{array}
\right)=
\left[ 
      \begin{array}{cc}
      T_{11} & T_{12}\\
      T_{21} & T_{22}\\
      \end{array}
\right] 
\left[ 
      \begin{array}{c}
      \hat{p}_{a} \\
      \hat{q}_{a} \\   
      \end{array}
\right]
\end{equation}

In Eq. \ref{T_matrix}, $\hat{p}_{a}$ and $\hat{p}_{b}$ are the acoustic pressures on the two sides of the sample and $\hat{q}_{a}$, $\hat{q}_{b}$ are the volume velocities at the same locations ($\hat{q}_{a} = \hat{u}_{a}A$, $\hat{q}_{b} = \hat{u}_{b}A$, being $A$ the sample area).

It can be easily shown that, under the assumption of constant particle velocity across the sample, $T_{11}=1$, $T_{21}=0$, $T_{22}=1$ and $T_{12}=(\hat{p}_1-\hat{p}_2)/q$, where $\hat{q}=\hat{q}_{a}=\hat{q}_{b}$. By multiplying $T_{12}$ by the sample area $A$, the transfer impedance is computed.
When a thin acoustic element is coupled with a back cavity, the transfer impedance of the panel and the cavity impedance $Z_{cav}$ form a series impedance $Z_{tot}=Z_{tr}+Z_{cav}$. The cavity impedance can be expressed as $Z_{cav}=-j \operatorname{cot} \left( w_l l \right)$, where $w_l$ is the wavelength and $l$ is the cavity length. Finally, the reflection and the absorption coefficients can be computed as follows: 

\begin{equation}\label{Refl_coeff}
\Re= \frac{Z_{tot}-1}{Z_{tot}+1}
\end{equation} 

\begin{equation}\label{alpha}
\alpha=1-\mid \Re \mid ^2
\end{equation}

\subsection{Determination of the particle velocity}\label{velocity}

During the tests performed for the present study, the transfer function $H_{ei}=p_i/e$ between the acoustic pressure $p_i$ at the microphone $i$ and the loudspeaker electric signal $e$ is recorded. As a consequence, through the Eq.\ref{u_test-sect} it is possible to carry out the transfer function of the particle velocity at the panel inlet cross section:

\begin{equation}\label{Hev}
H_{ev}= \frac{u_0}{e}= \frac{1}{\rho_0 c} \left[ H_{e+}\exp(\mathrm{i}k_0L) - H_{e-} \exp(-\mathrm{i}k_0L)\right] 
\end{equation}

In the Eq. \ref{Hev}, $H_{e+}$ and $H_{e+}$ are the measurable quantities 
\begin{equation}\label{H_e+}
H_{e+}= \frac{H_{e1}\exp(\mathrm{i}k_0x_2)-H_{e2}}   
                   {2\mathrm{i} \sin (k_0x_2)}  
\end{equation}
\begin{equation}\label{H_e-}
H_{e+}= \frac{H_{e2}-H_{e1}\exp(-\mathrm{i}k_0s)}   
                   {2\mathrm{i} \sin (k_0x_2)}  
\end{equation}

In order to get the amplitude of the particle velocity, which is needed for the non-linear characterization of the samples. the auto spectrum of the voltage $e$, $S_{ee}$, can be used, being the auto-spectrum of a sine wave signal equal to $1/2$ of the squared amplitude of the signal. As a consequence, the absolute value of the particle velocity is  
\begin{equation}\label{abs_u}
\left| u_v \right| = \left|\frac{u_0}{e}\right|=\frac{ \left| H_{ev} \right| \sqrt{2S_{ee}}}{K} 
\end{equation}
where $K$ [V/Pa] is the calibration factor of the microphones, needed to convert the electric signals of the microphones into the corresponding pressure values.

\subsection{Test facilities}
\begin{figure}
     \centering
        {\includegraphics[scale=0.9]{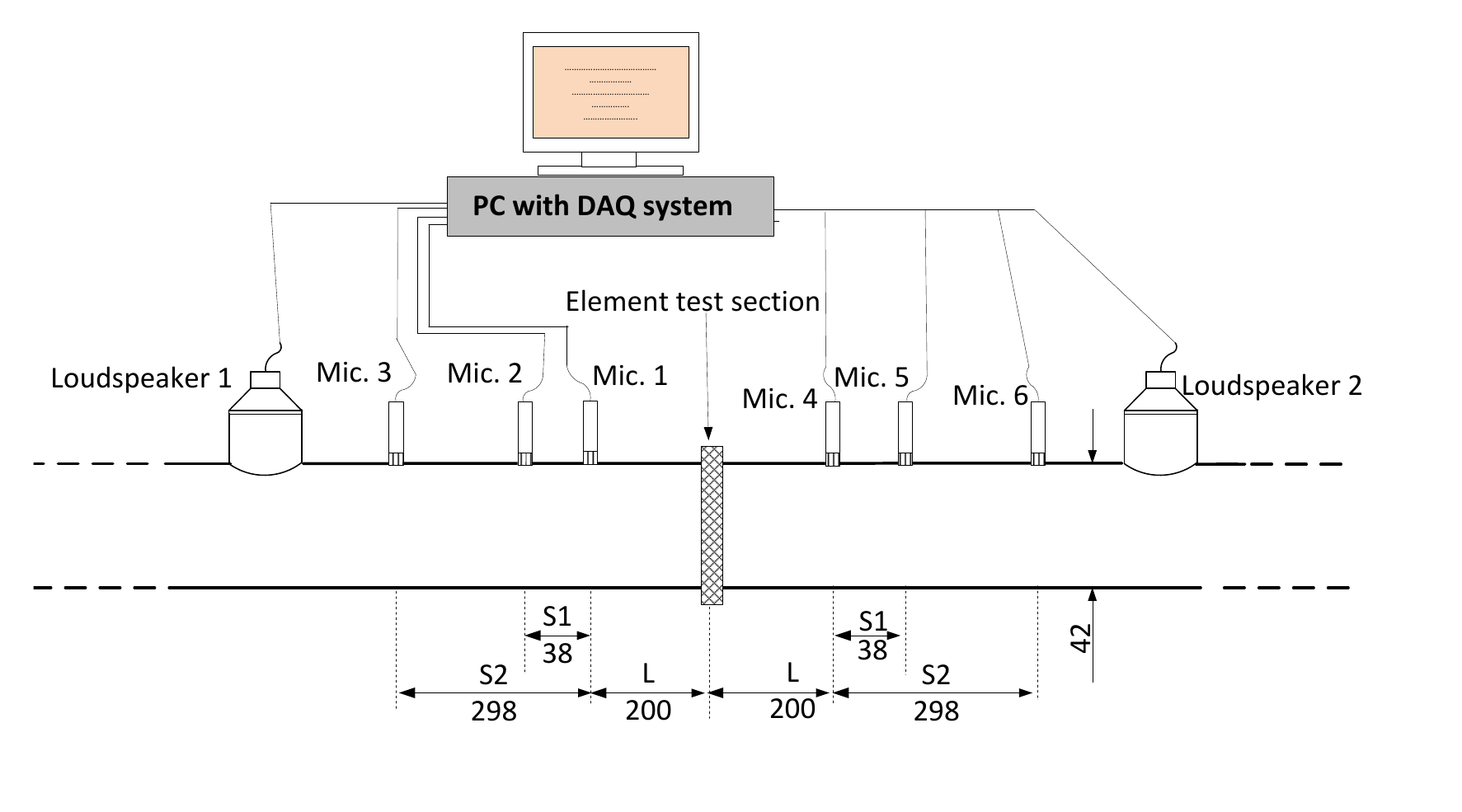}}
         \caption{Test-rig configuration}
         \label{fig_TestRig}
\end{figure}

The experimental work presented in this paper has been performed in the acoustic laboratory of Tallinn University of Technology (TUT).  In Figure \ref{fig_TestRig}, a schematic of the measurement set-up is depicted. 
All tests have been carried out in the absence of mean flow.
The measurement section has a round cross-section with the inner diameter $d=42mm$ and duct wall thickness $\delta =1.5mm$.
For the acoustical excitation, two professional series electrodynamic drivers (DAS$^{\circledR}$ ND-8) have been utilized. The drivers are mounted via side branches and driven by a software-based signal generator through National Instruments $^{\circledR}$ NI9269 analog output module and power amplifier (Velleman$^{\circledR}$ VPA2100MN). The signal acquisition has been performed by a dynamic signal analyser (National Instruments $^{\circledR}$ NI NIcDAQ 9174 and NI 9234), controlled by PC based virtual instrument (LabVIEW $^{\circledR}$).
The acoustic pressure has been measured by using six flush mounted $1/4$ inch prepolarized pressure microphones (type 40BD G.R.A.S.$^{\circledR}$, equipped with preamplifiers 26CB G.R.A.S.$^{\circledR}$). Two microphone separations $s_{1}=280$ mm and $s_{2}=35$ mm have been used. According to the relationship $0.1\pi< k_0 s_{i} < 0.8\pi$ ($i=1,2$) proposed in \cite{Abom_Error_analysis} for the two port experiments, the corresponding frequency region reckoned during the measurements ranged from $60$ Hz to $3.9$ kHz at temperature of $20 \ensuremath{^\circ}$ C.

\section{Description of the MGEs}\label{section_Description}

\subsection{General characteristics}

\begin{figure} [ht!]
     \begin{flushleft}
        {\label{fig_MGE_waves}
          \hspace{1cm}\textbf{a)} \includegraphics[scale=0.7]{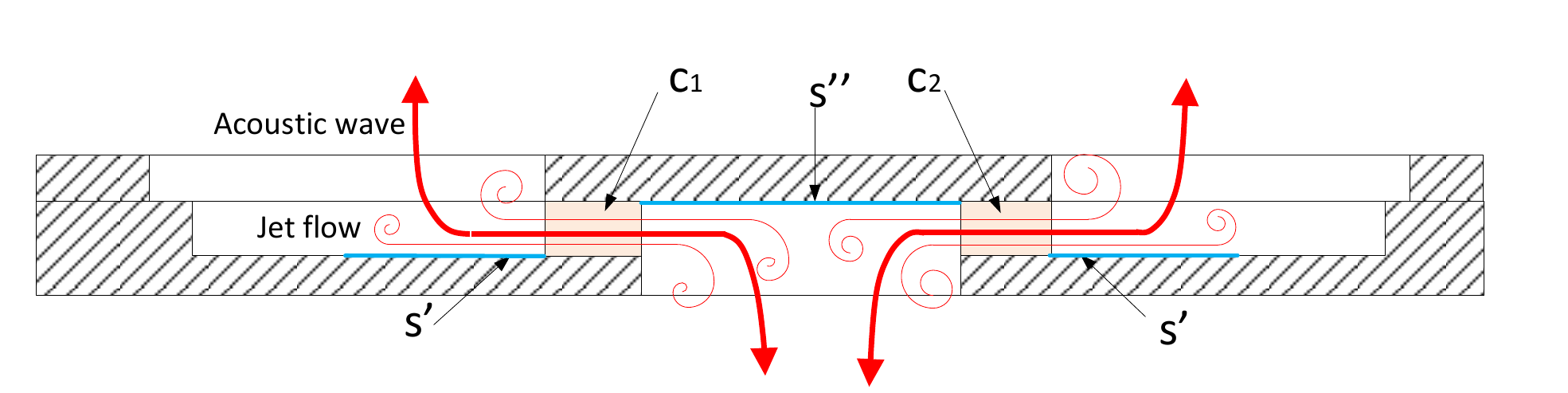}}  
        {\label{fig_MPE_waves}
        \\
          \hspace{2cm} \textbf{b)}\includegraphics[scale=0.7]{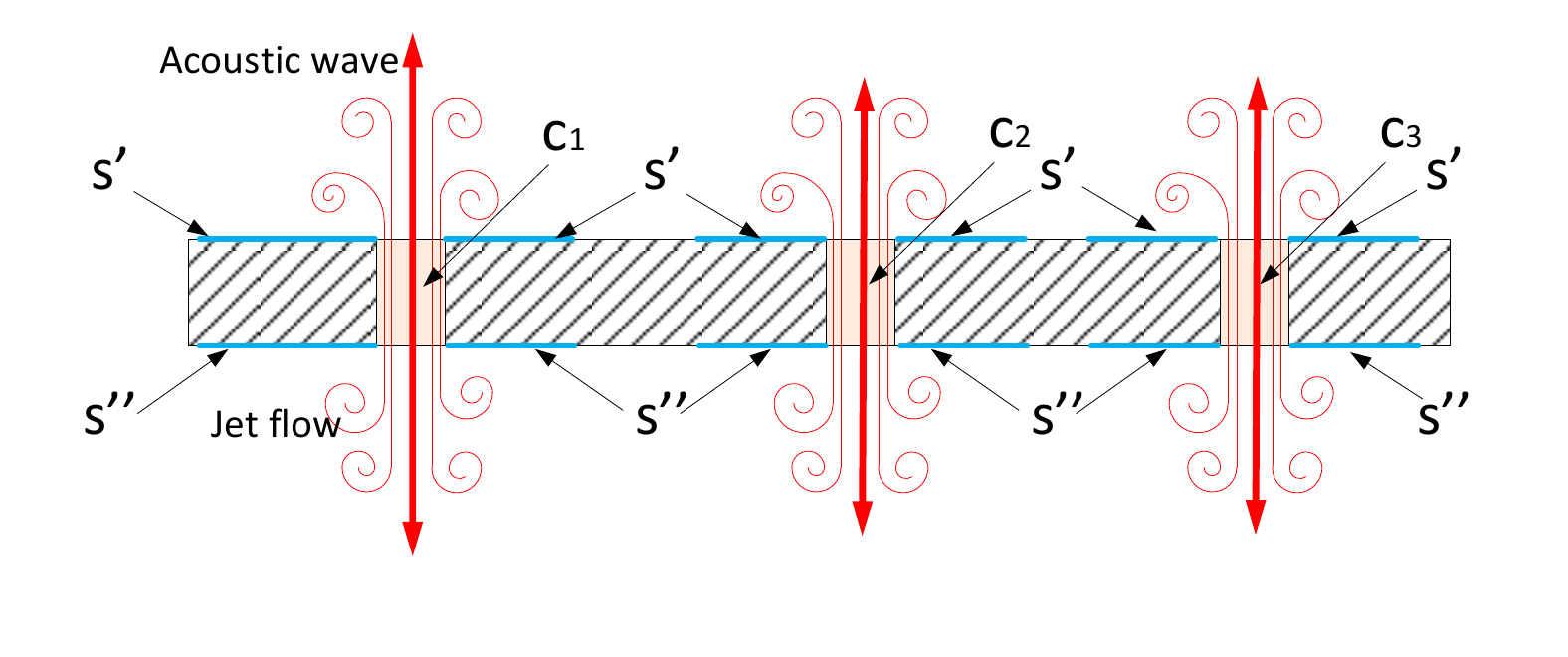}}   
      \end{flushleft}
    \caption{Propagation of the acoustic waves: a) in MGE; b) in MPE}  
   \label{fig_MGE_vs_MPE_waves}
\end{figure}

A MGE is a fibre-less acoustic absorbers constituted by two slotted layers and a number of internal micro-channels generated when the layers are mated. 
In the MGE the sound waves enter the element via the slots of the first layer \textit{(inlet)} and are forced to travel the network of micro-channels where the energy dissipation is primarily expected to take place. Eventually, the sound waves propagate out of the element via the slots at the second layer \textit{(outlet)} (see Fig. \ref{fig_MGE_vs_MPE_waves}a ).

\begin{figure}
     \begin{flushleft}
        {\label{fig_MGE_CADa}
          \hspace{0.1cm}\textbf{a)} \includegraphics[scale=0.80]{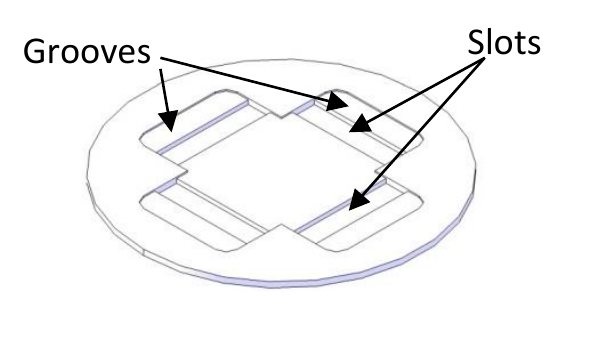}}  
        {\label{fig_MGE_CADb}
          \hspace{0.1cm} \textbf{b)}\includegraphics[scale=0.80]{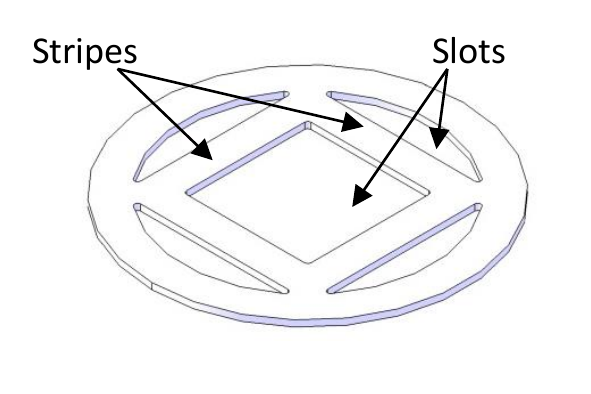}}   
          {\label{fig_MGE_CADc}
          \hspace{0.1cm} \textbf{c)}\includegraphics[scale=0.80]{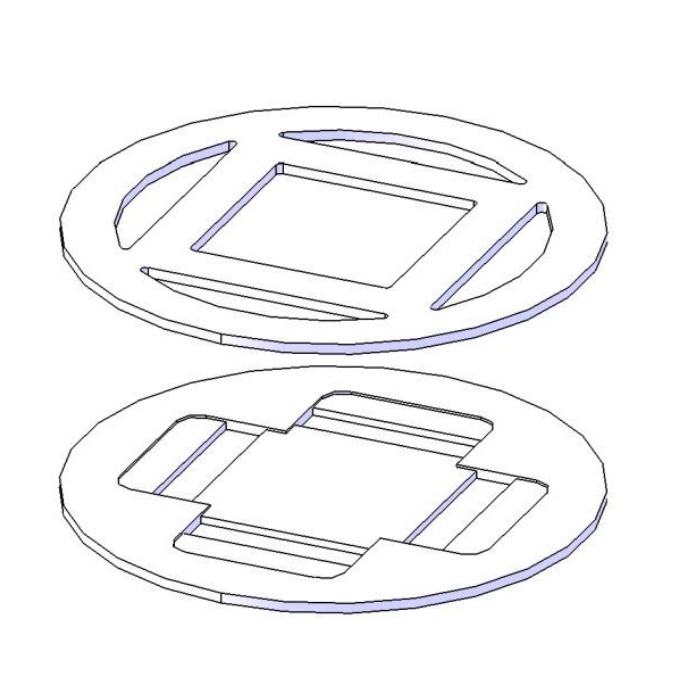}}  
      \end{flushleft}
    \caption{CAD model of the MGE: a) Outlet layer; b) Inlet layer; c) exploded view}  
         \label{fig_CAD_MGE}
\end{figure}

In the specific case of Fig. \ref{fig_CAD_MGE}, the inlet is a \textit{pseudo}-layer made of intersecting stripes, to contain the weight of the absorber, and the outlet is provided with grooves adjacent the slots. The internal micro-channels are generated as the stripes at the inlet partially overlap the grooves at the outlet. Punching and/or milling are the manufacturing processes involved in the production of MGEs.

In the configurations shown in Fig.s \ref{fig_CAD_MGE} and \ref{fig_MGE_vs_MPE_waves} a,  the micro-channels assume a well defined rectangular cross section. In Fig.\ref{fig_MGE_vs_MPE_waves} also the cross section of a MPE is shown in comparison. It is clear that the micro-channels and micro-perforations perform analogue functions in MGEs and MPEs respectively. 

In particular, the diameter of the micro-perforations is analogue to the depth of the micro-channels (both will be indicated s $d$). Moreover, it is straightforward to understand that the width of the stripes (at the inlet) and of the slots (at the outlet) determine the length of the micro-channels, $t$.

It will be shown in Sec.\ref{Linear behaviour}  that the geometrical features of the MGEs can be utilized to design simple acoustic elements with absorptive properties superior to MPEs with perforation diameters of $\approx 300-700 \mu m$.

The acoustic absorber presented in this paper shares several geometrical similarities with the double panel absorber presented by Randeberg in \cite{Randeberg_Perforated_absorbers}. However, in \cite{Randeberg_Perforated_absorbers}, the geometry of the layers and the position of the slots does not generate micro-channels as in MGEs and the dissipation of the energy occurs in the entire gap located between the two layers. As a consequence, the two advantages of the MGEs above mentioned have not been investigated in \cite{Randeberg_Perforated_absorbers} as well as the non-linear effects have been neglected.

A conclusive note in this paragraph is about the possibility of designing MGEs deprived of grooves, where the micro-channels are generated by simply spacing apart inlet and outlet layers which are not any more in contact. It is possible to use piezo-electric actuators which vary the gap between the layers according to the performance sought.

\subsection{Impedance decomposition}
According to Eq.\eqref{series_impedance_MGE}, the total transfer impedance of MGE is given by the transfer impedance of the layers and the one of the micro-channels. Nevertheless, experiments - not reported here for brevity - show that the contribution of the layers on the overall performance is practically negligible. In fact, the presence of wide slots implies nearly insignificant resistance and minor reactance, being the last one more meaningful as the thickness of the layers is larger than $1 mm$.  


However, this study is focussed on MGEs provided with layer thickness of $\approx 0.5 mm$, for which it is safe to write: $Z_{MGE} \approx Z_{Ch}$. 


\section{Linear behaviour} \label{Linear behaviour}
In the linear regime, the dissipation of the acoustic energy mainly occurs in the viscous layers within the micro-channels, due to the viscosity of the fluid and the presence of solid walls. The thickness of the oscillating viscous layers is expressed by the Stoke number, $\delta_v=\sqrt{\mu / \omega \rho_0}$, with $\mu$ is the air viscosity ($ \approx 1.85 x 10^{-5} kg/m s$) in standard conditions. In MPEs and MGEs the oscillating viscous layers extend as much as to cover almost completely or completely the cross section of the apertures. This condition can expressed as \textit{Sh=$\mathcal{O}$(1)}, where \textit{Sh} is the Shear number of the apertures:

\begin{equation}\label{Shear_number}
\textit{Sh}=\frac{d}{2 \delta_v}=d \sqrt{\frac{\omega \rho_0}{4 \mu}}
\end{equation} 

For MPEs and MGEs, $d$ in Eq. \ref{Shear_number} is the diameter of the perforations and the depth of the micro-channels respectively. For low enough excitation amplitudes, the amplitude of the particle velocity frequency components ($\hat{u}$) and the amplitude of the acoustic pressure frequency component ($\hat{p}$) are linearly dependent and the transfer impedance is calculated as a function of the perforation geometry and frequency only. This is the case treated in this paragraph.
  
The linear transfer impedance can be seen as $Z_{Lin}=Z_{int}+Z_{ext}$, where $Z_{int}$ is the \textit{internal part}, and $Z_{ext}$ the \textit{external part} of the impedance, due to phenomena which take place within and outside the apertures respectively.
\subsection{Internal part}
In case of MPEs, both contributions have been modelled by Crandall first \cite{Crandall_Theory}, and Maa afterwards \cite{Maa5_Basic_theory_of_acoustics}. However, here the formulation proposed by Allard, for the internal impedance of slit shaped apertures is recalled  \cite{Allard_book}, since it fits well to the geometry of the micro-channels in MGEs: 

\begin{equation}\label{z_int}
z_{int}=\frac{j\omega t}{\sigma c}\left[ 1-\frac{\tanh\left( \delta_v \sqrt{j}\right) }{\delta_v\sqrt{j}} \right]
\end{equation}
\\
where $t$ is the length of the micro-channels and $j$ the imaginary unit. In general, for the small apertures or low frequencies, the loss mechanism described by  Eq.\eqref{z_int} is frequency independent. 
\subsection{External part}
\begin{figure}
     \begin{flushleft}
        {\label{fig_End_Res_Extraction}
          \hspace{0.1cm}\textbf{a)} \includegraphics[scale=0.31]{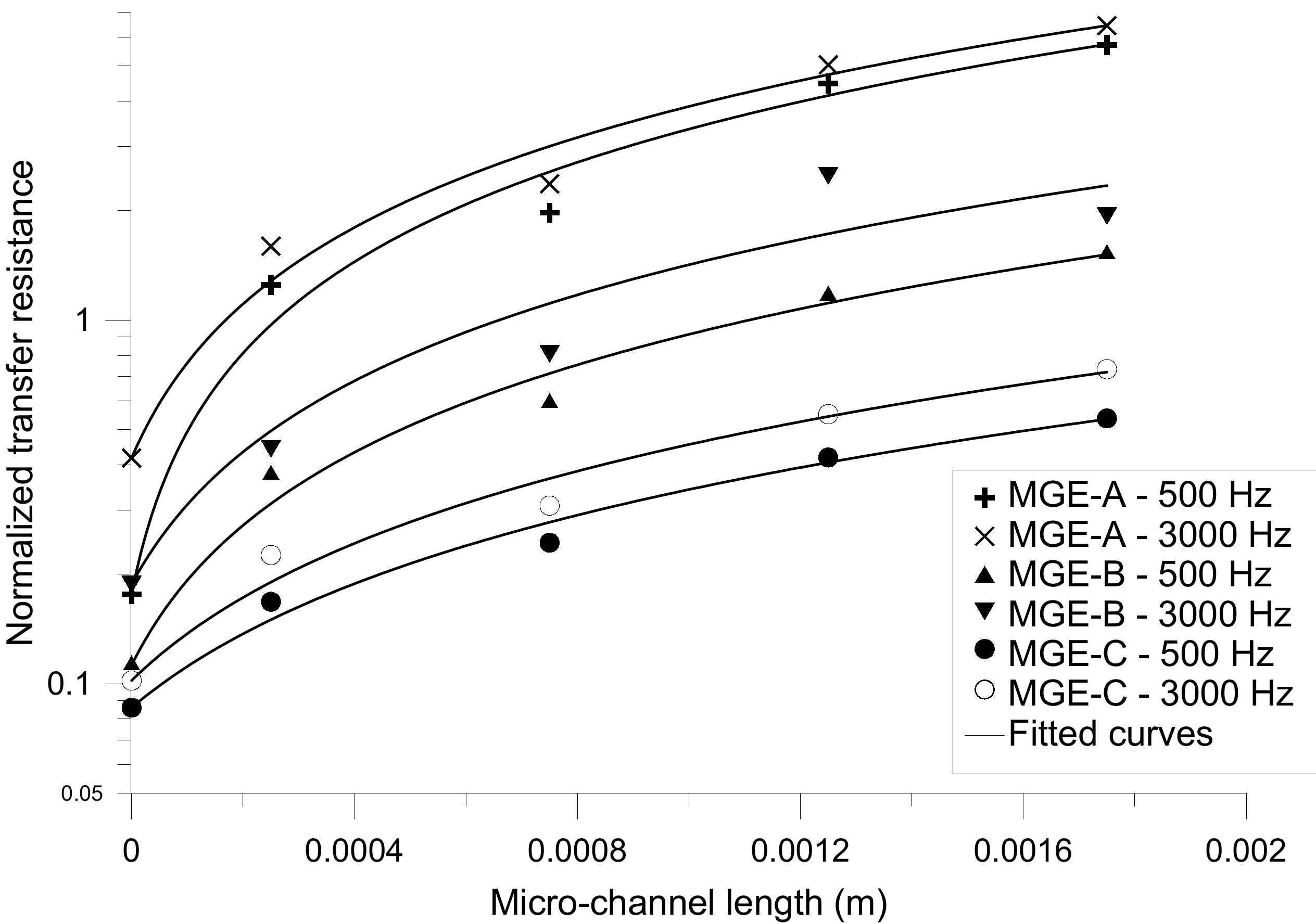}}  
        {\label{fig_End_Rea_Extraction}
          \hspace{0.1cm} \textbf{b)}\includegraphics[scale=0.31]{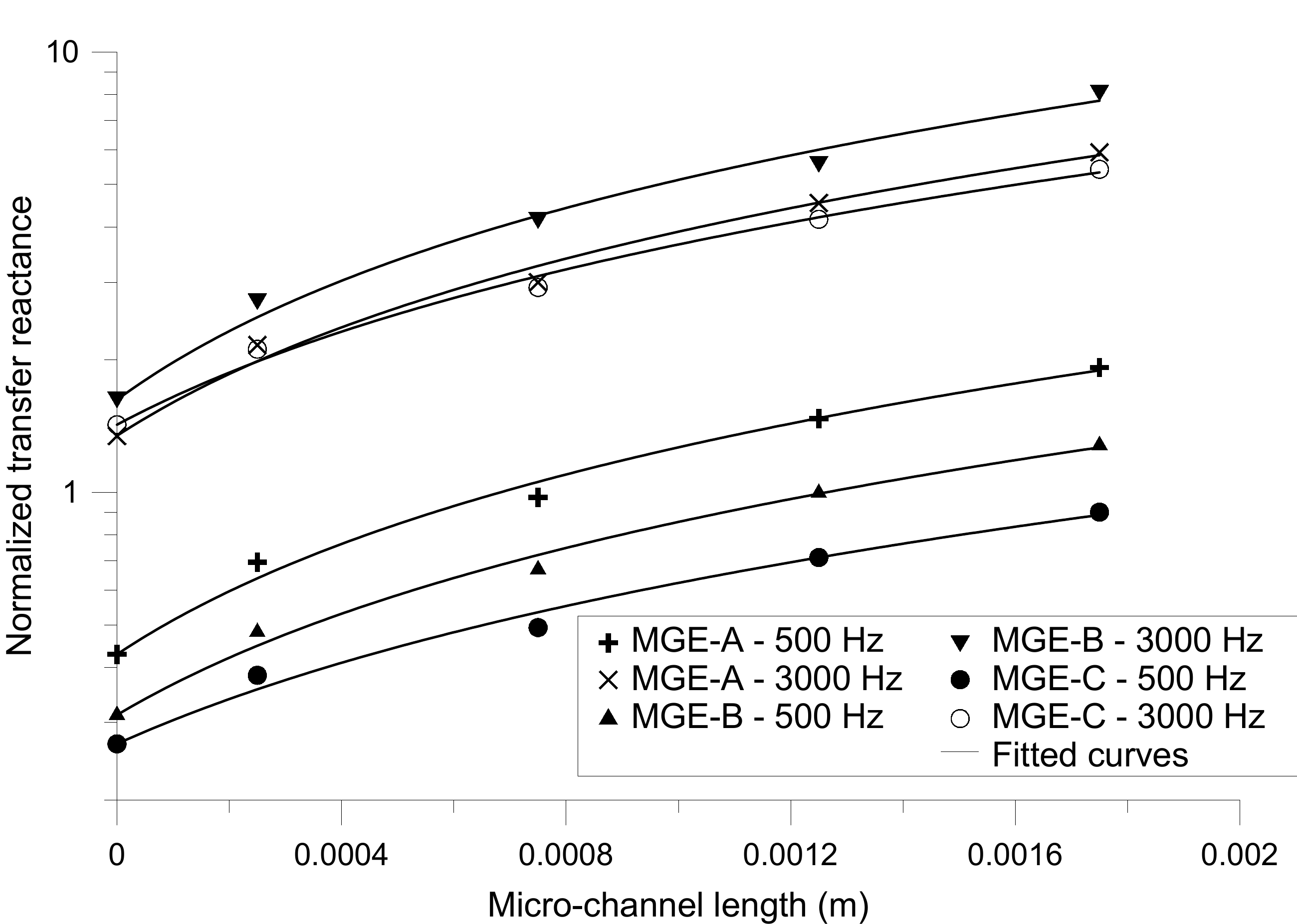}}   
      \end{flushleft}
    \caption{Measurement of the end effects: a) Resistive part; b) Reactive part}  
         \label{fig_End_Res/Rea_Extraction}
\end{figure}

The motion of the air outside the apertures affects both the dissipative and inertial part of the overall transfer impedance, thus end-corrections for both the real and the imaginary parts of the transfer impedance must be introduced. From the experimental point of view a methodology which particularly fits to MGEs has been utilized to measure the external part of the transfer impedance. 
It consists of measuring the transfer impedance for several samples with same cross section but different length of the apertures, which is done by simply using different inlet layers (i.e. different $q$ values, see Figure \ref{fig_CAD_MGE}). By linearly interpolating the values of transfer impedance, the normalized end corrections can be extrapolated at the channel length $t=0$. In this case, the only contribution to the aperture impedance is given by the end-corrections.   
This approach is purely experimental and it is based on the assumption (confirmed by \cite{Temiz_End_corrections} in case of sharp edged circular perforations with $t/d \geqslant 0.5$), that the length of the apertures does not affect the end-corrections. 
Figure \ref{fig_End_Res/Rea_Extraction} exhibits the resistive and reactive values of the micro-channels measured at $500$ and $3000$ Hz. The samples have channel lengths varying between $2.50$E-04 m to $2.75$E-03 m ands amples with the same cross-section are denominated by the same name of MGE (e.g. MGE-A, MG-B,...). The procedure is repeated at every frequency of interest.  

\begin{figure}
     \begin{flushleft}
        {\label{fig_End_Resistance_vs_Shear}
          \hspace{0.1cm}\textbf{a)} \includegraphics[scale=0.31]{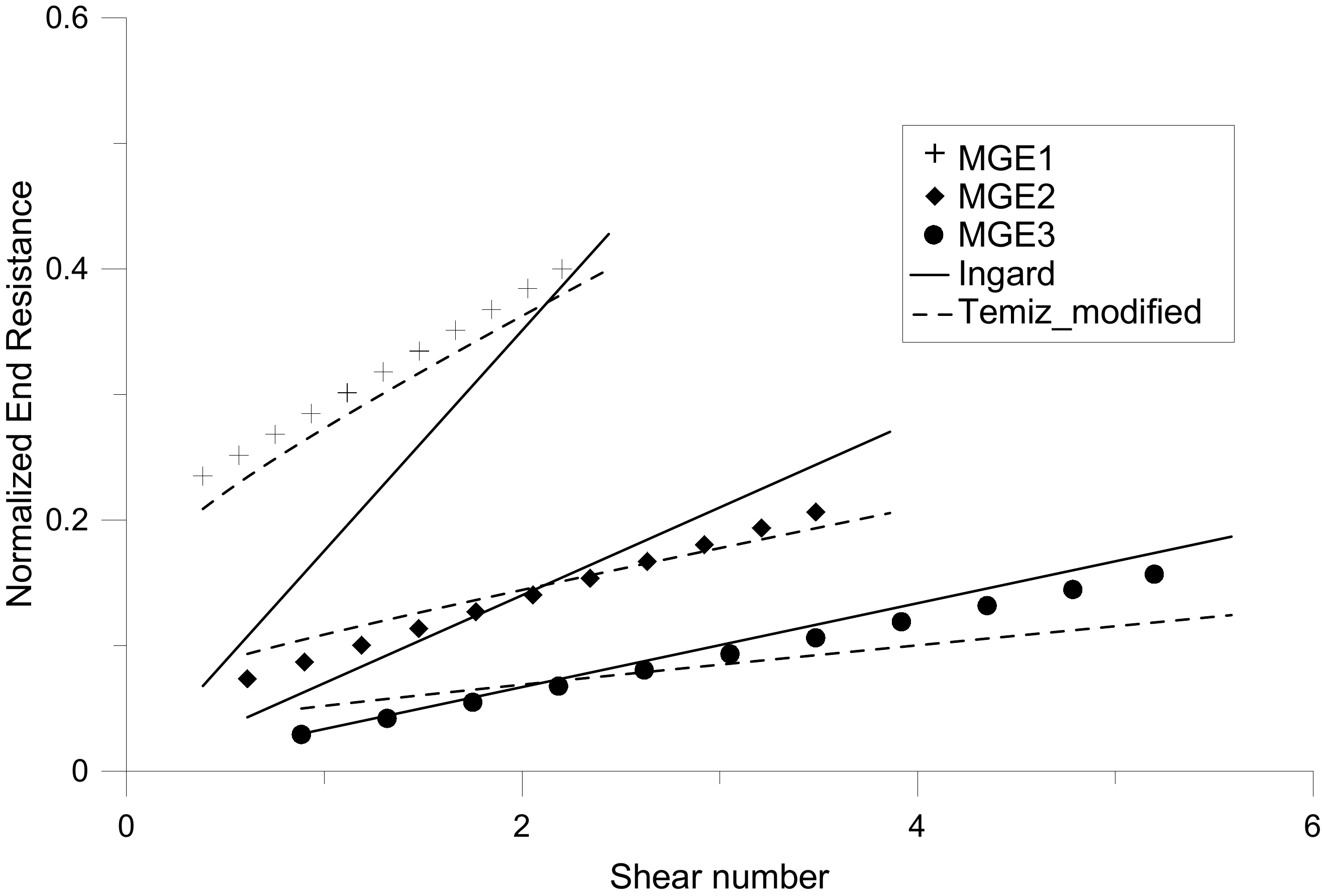}}  
        {\label{fig_End_Reactance_vs_Shear}
          \hspace{0.1cm} \textbf{b)}\includegraphics[scale=0.30]{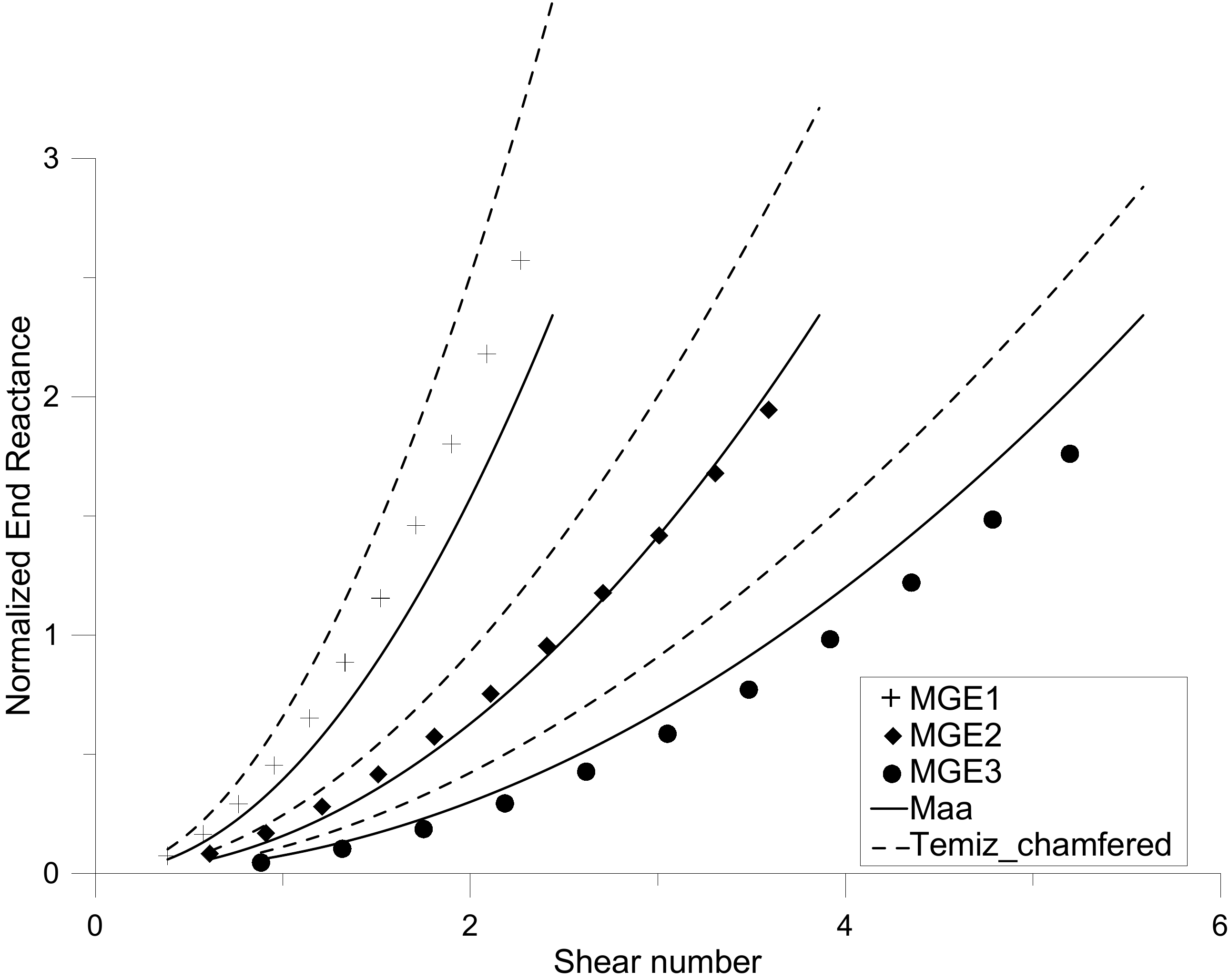}}   
      \end{flushleft}
    \caption{Experimental and theoretical values of the external transfer impedance: a) End-resistance; b) End-reactance}  
         \label{fig_End_Res/Rea_vs_Shear}
\end{figure}

Ingard \cite{Ingard_Acoustic_resonators} estimated the resistive end correction of a circular duct as a function of the power dissipated by an oscillating motion of a fluid on a plane surface
\begin{equation} \label{Ingard_Rs}
R_{s}=\frac{1}{2}\sqrt{2\nu\rho_0\omega}
\end{equation}
For two ends of a narrow tube the total resistance is $2R_s$, hereby the end resistive correction proposed by Maa for slit shaped holes \cite{Maa5_Basic_theory_of_acoustics} is:
\begin{equation} \label{r_MPE_e}
r_{ext}=\frac{2 \alpha R_{s}}{\sigma\rho_0 c}
\end{equation}
where $\alpha=2,4$ for rounded and sharp shaped apertures respectively \cite{Allam_A_new_type_of_muff}. Temiz investigated the dependence of the end corrections on the edge configuration \cite{Temiz_End_corrections} by using the following non-dimensional parameter to generalize the results for the resistive part: 
\begin{equation} \label{Temiz_alpha}
\alpha_{T}=\frac{ X_{ext}  \sigma }{2 R_s}
\end{equation}
In Eq. \ref{Temiz_alpha} $X_{ext}=\chi_{ext} \rho_0 c=\Re \left( Z_{ext}\right)$. In case of MGE samples examined here, the edges of the micro-channels are characterized by the presence of $13 \mu m$ chamfer (see Section \ref{Non-Linear behaviour}).
In \cite{Temiz_End_corrections}, $\alpha_{T}$ is given for circular apertures with different edge geometries, obtained by curve fitting experimental and numerical data of perforated plates with large perforation diameter (4.2 mm). These expression should be general and applicable to apertures with different values of diameters. However, when applied to apertures with sub-millimetre diameter, the resistive end-correction shows a minimum at values of $\textit{Sh}=\mathcal{O} \left( 1 \right)$, which limits the applicability of these formulas to apertures with over-millimetre diameter. By assuming that the cross section geometry of the apertures has a minor influence on the resistive end corrections, the expression of $\alpha_{T}$ given by Temiz in \cite{Temiz_Nonlinear_acou_transfer} for apertures with chamfered edges is here proposed in a modified version by using the experimental data of MGEs : 
\begin{equation} \label{Temiz_alpha_chamfered}
\alpha_{cf}=c_1 \textit{Sh}^{c_2}+1.70+1.1c^{*1.74} \textit{Sh}^{-0.26}
\end{equation}
where $c_1=5.08$, $c_2=-1.45$ for apertures with millimetre apertures (as stated in \cite{Temiz_Nonlinear_acou_transfer}) and $c_1=4.05$, $c_2=-0.90$ for sub-millimetre apertures, as in case of MGEs. In Eq. \ref{Temiz_alpha_chamfered} $c^*=c_p/d$ where $c_p$ is the chamfer length.
In case of MGEs, the correct expression for $\alpha$ is:
\begin{equation} \label{alpha_MGE}
\alpha_{MGE}=\frac{1}{2} \left[ 4.05 \textit{Sh}^{-0.90}+1.70+1.1c^{*1.74} \textit{Sh}^{-0.26} \right]
\end{equation}
In Fig. \ref{fig_End_Res/Rea_vs_Shear}a the experimental results are compared with the results obtained by using Eq.\ref{alpha_MGE} (\textit{"Temiz\_ modified"}) and the ones obtained from Eq.\ref{r_MPE_e} with $\alpha=2$ (\textit{"Ingard"}). Results from Eq.\ref{r_MPE_e} present a noticeable discrepancy with the experimental ones only in case of MGE1, where $d=120 \mu m$ and a few $\mu m$ error in the measurement of $d$ could have affected the predictions. 

The choice $\alpha=2$ in Eq. \ref{r_MPE_e} could lead to the conclusion that the edges of the micro-channels are rounded. However, due to the the small value of the chamfered length, it seems to be more reasonable to consider the  edges as almost sharp and the presence of the surfaces $s ^{\prime}$ and $s ^{\prime \prime}$ (see Fig. \ref{fig_MGE_vs_MPE_waves} a)) halves the resistive value of the end-corrections, by preventing the energy dissipation nearby. This is highlighted by the presence of the coefficient 1/2 in the Eq. \ref{alpha_MGE}. 
The vibrating air plug within the aperture behaves as the acoustic mass of an aperture whose actual axial dimension is given by the thickness of the aperture $t$ plus the reactive end-correction lengths on both sides of the panel \cite{Ingard_Acoustic_resonators}. 

For slit-shaped apertures Maa \cite{Maa5_Basic_theory_of_acoustics} suggests that the rectangular cross section can be modelled as an equivalent ellipse. By introducing an $ellipticity$ factor
\begin{equation} \label{ellipticity}
\overline{e}=\sqrt{1-\left( \frac{d}{w}\right) ^{2}}
\end{equation} 
where $d$ and $w$ are the height and the length of the slit, i.e. the depth and the width of the micro-channels in case of MGEs, the reactive impedance is written as 
\begin{equation} \label{chi_ext}
\chi_{slit-ext}=\frac{1}{2}\omega\rho_0 dF\left(\overline{e}\right), 
\end{equation}
where 
\begin{equation} \label{F-e}
F\left( \overline{e}\right) =\int_{0}^{\pi/2} \frac{d\theta}{\left({1-\overline{e}^{2}\cos\theta}\right)^{2}}
\end{equation}
For an entire MGE, the normalized external part of the transfer reactance becomes:
\begin{equation}\label{Chi_ext_panel}
\chi_{ext}=\frac{\omega d F\left(\overline{e}\right)}{2 \sigma c}
\end{equation}
%


In \cite{Temiz_End_corrections} Temiz provided a formulation for reactive end-corrections of apertures with different edge geometries. This formulation is based on the definition of the following parameter, which has been used to generalize the results:

\begin{equation} \label{Temiz_delta}
\delta_{T}=\frac{ X_{ext} \sigma }{\rho_0 \omega d/2}
\end{equation}

where $X_{ext}=\Im \left( Z_{ext}\right)$. The following expression is given in \cite{Temiz_End_corrections} for circular apertures with chamfered edges:

\begin{equation} \label{Temiz_delta_chamfered}
\delta_{T-ch}=0.97 \exp \left (-0.20 \textit{Sh} \right)+1.54+0.97 c^{*0.56} \exp \left (-0.01 \textit{Sh} \right)
\end{equation}

For micro-channels the following equation fits well: 
\begin{equation} \label{delta_MGE}
\delta_{MGE}=F\left( \overline{e} \right) \delta_{T-ch} 
\end{equation}
 
Fig. \ref{fig_End_Res/Rea_vs_Shear}b shows a good match between the experimental results, the results obtained by modifying the formulation by Temiz (Eq. \ref{delta_MGE}), and the one provided by Maa for slit shaped apertures (Eq. \ref{Chi_ext_panel}). It means that the presence of the surfaces $s ^{\prime}$ and $s ^{\prime \prime}$ (see Fig. \ref{fig_MGE_vs_MPE_waves} a)) does not influence the inertia of the vibrating air plugs and the air in proximity of the apertures is still free to move as in the case of MPEs (Fig. \ref{fig_MGE_vs_MPE_waves}). 
\subsection{Absorption coefficient}
\begin{figure}
     \begin{flushleft}
        {\label{fig_MGE_vs_MPE_linear_impedance}
          \hspace{0.1cm}\textbf{a)} \includegraphics[scale=0.31]{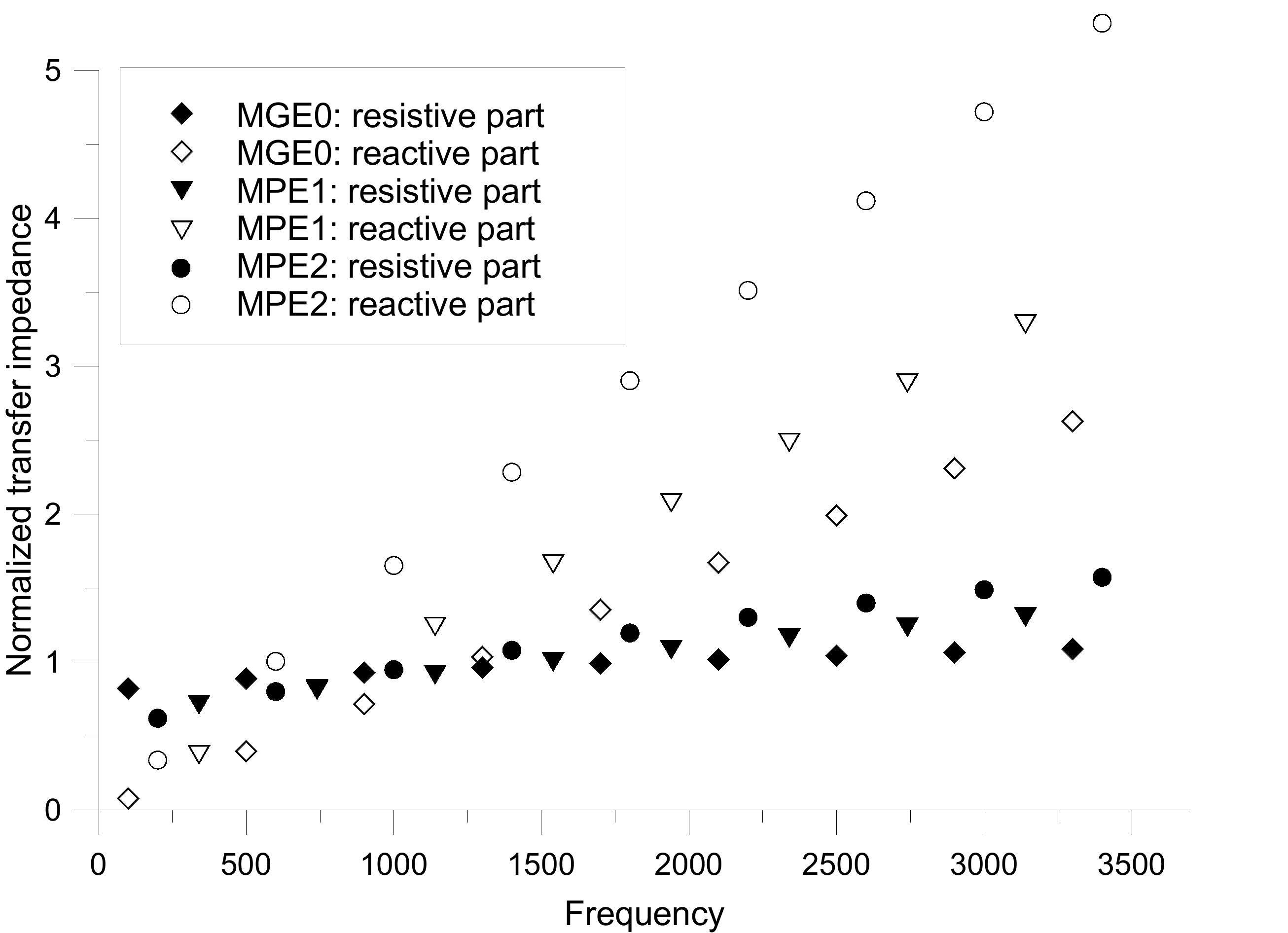}}  
        {\label{fig_MGE_vs_MPE_linear_alpha}
          \hspace{0.1cm} \textbf{b)}\includegraphics[scale=0.31]{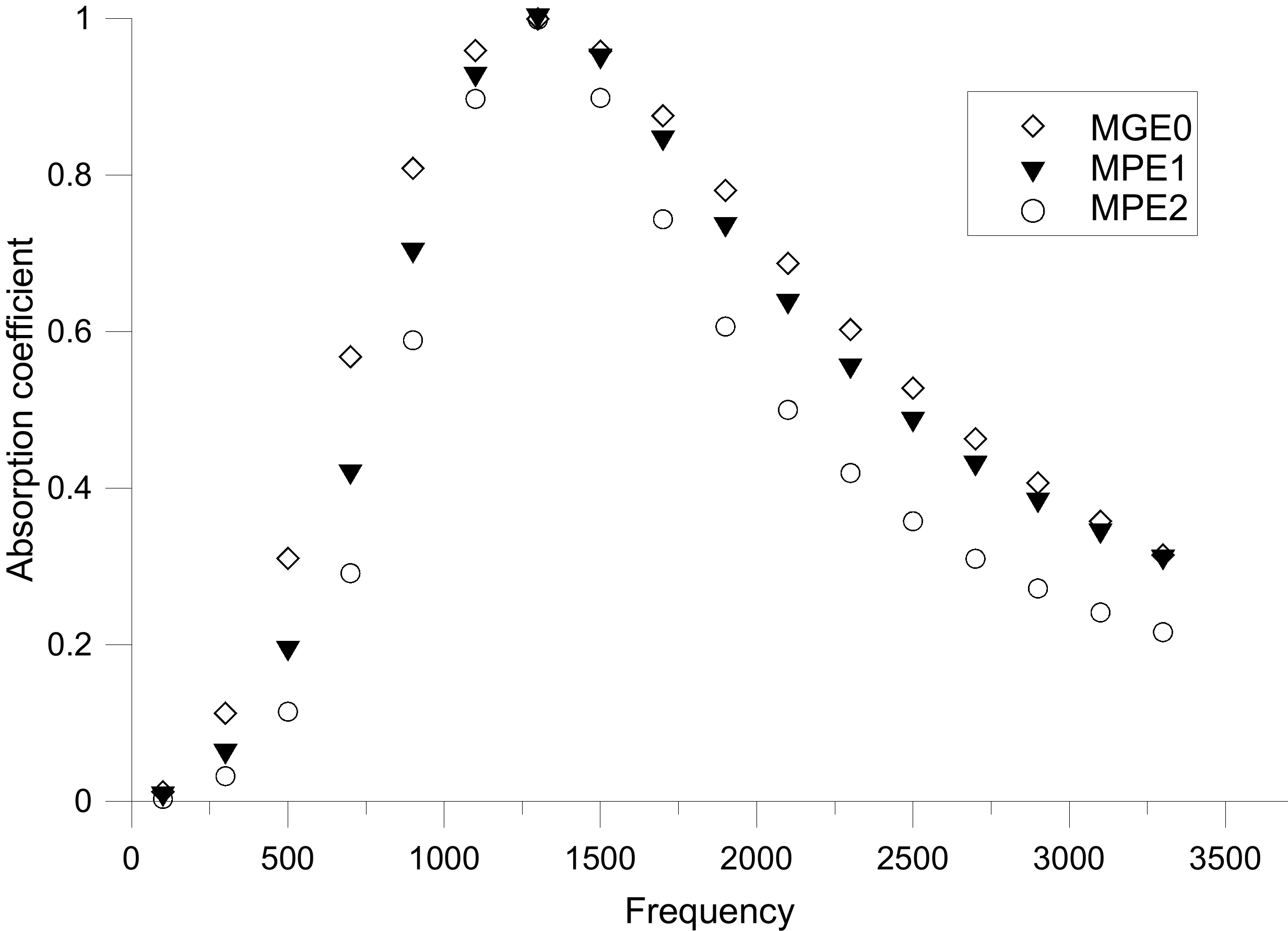}}   
      \end{flushleft}
    \caption{Linear performance of MGE and MPE samples: a) Resistive and reactive normalized tranfer impedance; b) Absorption coefficients.}  
         \label{fig_MGE_vs_MPE_linear}
\end{figure}

In Fig. \ref{fig_MGE_vs_MPE_linear} the acoustic performance of different micro-grooved and micro-perforated samples is shown in terms of normalized transfer impedance and absorption coefficients. The geometrical characteristics of these samples are listed in Tab. \ref{Tab_1}. The normalized resistive transfer impedance of MGE0 is unitary since very the low frequencies and the trend is basically flat. The reactive part of the MGE0, remains lower than the recative parts of the MPE1 and MPE2 at every frequency. As a result, the absorption coefficient of  MGE0 is $\approx 31\%$ higher than the one of the MPE2 and $\approx 10\%$ higher than the one of the MPE1. 
 
It must be pointed out that typical MPEs have diameters $\approx 300-700 \mu m$, porosities $\approx 1$ and thickness $t \approx 0.5 mm$. Aperture diameters smaller than $\approx 300-400 \mu m$ accompanied by higher porosities can be technologically demanding, as for the case described in \cite{Quian_Ultra_micro_perforations}. From this point of view the MPE1 can be considered already as less cost-effective than typical MPEs because of the reduced diameter of the perforations and the density of $\approx 22$ holes per square centimetre necessary to provide porosity $\sigma=1.53\%$.  Moreover, thickness smaller than $\approx 400-500 \mu m$ can hinder the perforation process other than compromise the mechanical properties of the micro-perforated plate.
In case of MGEs, the mechanical properties are guaranteed by the thickness of the entire element ($1 mm$ for the samples examined here). Furthermore, by using simple technological processes such as milling and/or punching, it is possible to manufacture micro-channels with  small depths, down to  $\approx 100 \mu m$, short lengths ($100-250 \mu m$) and $\approx 3$ micro-channles per square centimetre, as for MGE0. 
This represents a \textit{double advantage} from the acoustical point of view and explains the favourable trends shown in Fig. \ref{fig_MGE_vs_MPE_linear}. In fact, as known from the literature \cite{Maa5_Basic_theory_of_acoustics}, a flat unitary broadband absorption coefficient is theoretically obtained when the normalized real part of the transfer impedance (resistance) is equal to $1$ and the normalized imaginary part (reactance) is null. It is possible to tend to such an ideal condition when the diameter of the micro-perforations (in MPEs) or the depth of the micro-channels (in MGEs) decrease and the porosity is increased adequately. From this point of view, the small depth of the micro-channels is definitely an \textit{advantage}. 

However, absorbers with short length of the apertures require lower porosities than absorbers with longer lengths in order to obtain the same performance. This fact represents the \textit{second advantage} and it can be explained by considering that elements with shorter apertures typically have lower reactance and resistance, as the other geometrical parameters are kept constant. Nevertheless, in this case resistance, reactance and absorption coefficient are recovered by simply reducing the porosity. As a consequence, notwithstanding the small depth of the micro-channels of MGE0, a relatively low value for the porosity ($\sigma=1.13$) is enough to guarantee adequate acoustic performance of the absorber. 

It will be shown in the following section that the presence of the surface $s'$ and $s''$ ahead and at the bottom of the micro-channels respectively (see Fig. \ref{fig_MGE_vs_MPE_waves} a), remarkably affects the end-corrections and the non-linear behaviour of the absorber, resulting in a reduced resistance - compared to the case of MPEs - at high levels of sound excitations. 

\section{Non-Linear behaviour} \label{Non-Linear behaviour}
The non linear behaviour of MGEs presents unique aspects due to the internal geometry of these elements.
In order to analyse it, several samples have been tested in presence of different levels of sound excitation. As known from literature, this behaviour occurs for Strouhal numbers $\textit{Sr}\ll 1$, i.e. when the particle displacement is much larger than the aperture size. In these conditions, and for circular apertures, Cummings and Eversman proposed the following relationship \cite{Cummings_High_amplitude_transmission}:
\begin{equation}\label{Cumming_non_lin_formula}
|\hat{p}_a|\simeq\rho_0 |\hat{u}_p|^2 \frac{1-\sigma^2 C_{vc}^2}{2 C_{vc}^2}
\end{equation}
where $C_{vc}=\dot{m} / \left( A \sqrt{2 \rho_0 \Delta p} \right) $ is the discharge coefficient of the orifice, related to the \textit{vena contracta}, i.e. the reduction of the diameter of the flow stream within the orifice. 
However, the transition form the linear regime, which occurs for \textit{Sh=$\mathcal{O}$(1)}, is not sudden, but an intermediate, almost linear regime can be detected as $\textit{Sr} > 1$. In this case, Disselhorst et al. presented in \cite{Disselhorst_Almost_linear_regime} a formulation valid for high Shear numbers (thus, not applicable to micro-aperures, where $\textit{Sh}=\mathcal{O}\left(1 \right)$) and in \cite{Temiz_Nonlinear_acou_transfer} Temiz presented a semi-empirical approach for MPEs which allows expressing the resistance and reactance changes of MPE, due to the non-linear effects, as a function of Shear and Strouhal numbers within the apertures. This approach is based on the following normalizations:
\begin{equation}\label{Temiz_NL_res}
\overline{R}_{NL}=2 C_{vc}^2\frac{ R_{NL}\sigma}  {|\hat{u}| \rho_0}
\end{equation}

\begin{equation}\label{Temiz_NL_rea}
\overline{X}_{NL}=\frac{ X_{NL} \sigma}  {\omega \rho_0 d/2}
\end{equation}

where $R_{NL}=\left[\Re\left( Z \right)-\Re \left( Z_{Lin} \right) \right]$ and $X_{NL}=\left[\Im\left( Z \right)-\Im \left( Z_{Lin} \right) \right]$ being $Z$ the acoustic impedance of the samples in quasi-linear and non-linear regimes.

In the present investigation, the same methodology is applied to the MGEs and, by curve fitting the experimental results normalized as in Eq.s \ref{Temiz_NL_res} and \ref{Temiz_NL_rea}, a generalized expression of the transfer impedance of the MGEs for the non-linear and quasi-linear regimes is provided.
To this aim it is necessary to have a good estimation of the discharge coefficient. $C_{vc}$ for sharp edged circular perforations is $\approx 0.611$, as shown in \cite{Cummings_High_amplitude_transmission}. However, this value changes for rectangular slits, is influenced by the porosity of the panel and is strongly affected by the geometry of the edges \cite{Hofmans_Aeroacoustic_resp_of_slit_quasi_steady}. In fact, rounding edges by a few percent of the aperture size can reduce the vena contacta almost completely and result in a higher $C_{vc}$. The edges of the micro-channels in the MGEs have been observed with a 100x magnification and the chamfer length has been found to be $\approx 13 \mu m$ in all samples. 
\subsection{Discharge coefficient}
The discharge coefficients of the rectangular micro-channels in MGEs have been computed by performing a steady state CFD simulation where the pressure at inlet and at the outlet were imposed. The steady state assumption obviously represents an approximation because the actual flow through the orifice is alternating during successive acoustic half-cycles.

\begin{figure}
        {\label{fig_CFD_mesh_MGE}
          \textbf{a)}\includegraphics[scale=0.8]{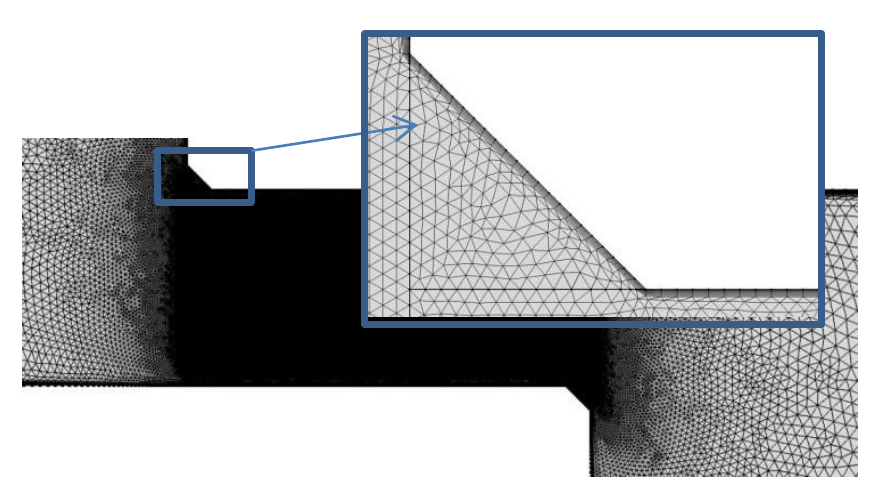}}  
         {\label{fig_CFD_velocity_MGE}
          \textbf{b)}\includegraphics[scale=1]{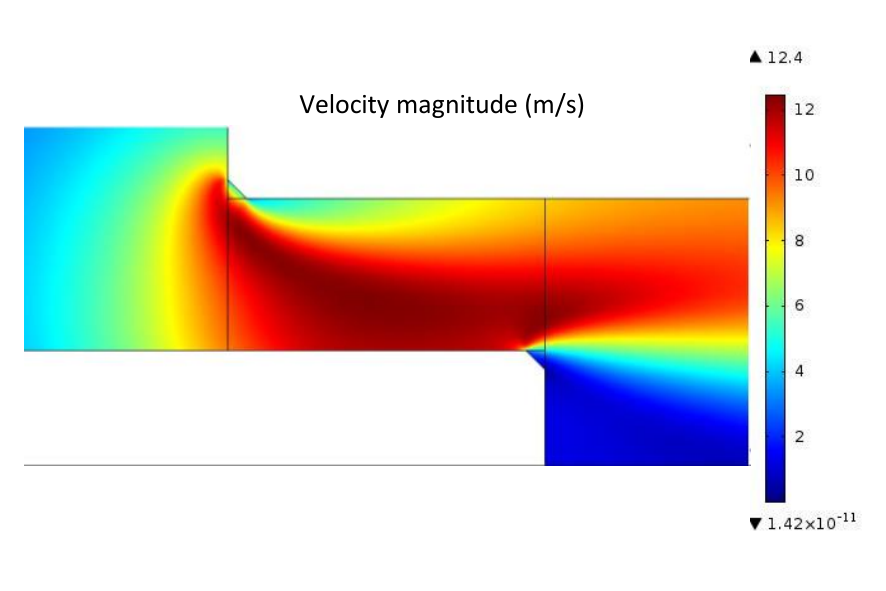}}  
\caption{CFD simulation of a micro-channel with chamfered edges: a) Mesh; b) Velocity profile.}  
         \label{fig_CFD_MGE}
\end{figure}

The solver utilized is COMSOL Multiphysics $^{\circledR}$ and the meshes, as shown in Fig. \ref{fig_CFD_MGE}, 
are obtained by a combination of triangular and rectangular elements, being the second ones mainly present in the vicinity of the boundaries with no-slip conditions. The simulations have been repeated by using the geometries of the micro-channels of all the samples studied here. It has been found that, for a MGE with $\sigma=1.14$, $d=120 \mu m$  and by imposing $1 mbar$ pressure at inlet, the the discharge coefficient varies from $C_{vc}=0.68$ to $C_{vc}=0.79$ by varying the geometries of the edges from  sharp to chamfered with $13 \mu m$ chamfer length.
Moreover, the MGEs present a $C_{vc}$ which is typically $\approx 10\%$ always higher than the $C_{vc}$ of a MPE with same $d$ and porosity. With a good approximation, the discharged coefficient of the MGE samples studied here, in the conditions examined in this paper, can be considered $C_{vc}\approx 0.8$. 
For the MPEs tested here, the averaged value of the discharge coefficient computed with the CFD simulation is $C_{vc}\approx 0.7$.

\begin{figure}
\begin{flushleft}
        {\label{fig_NL_res_MGE1}
          \textbf{a)}\includegraphics[scale=0.3]{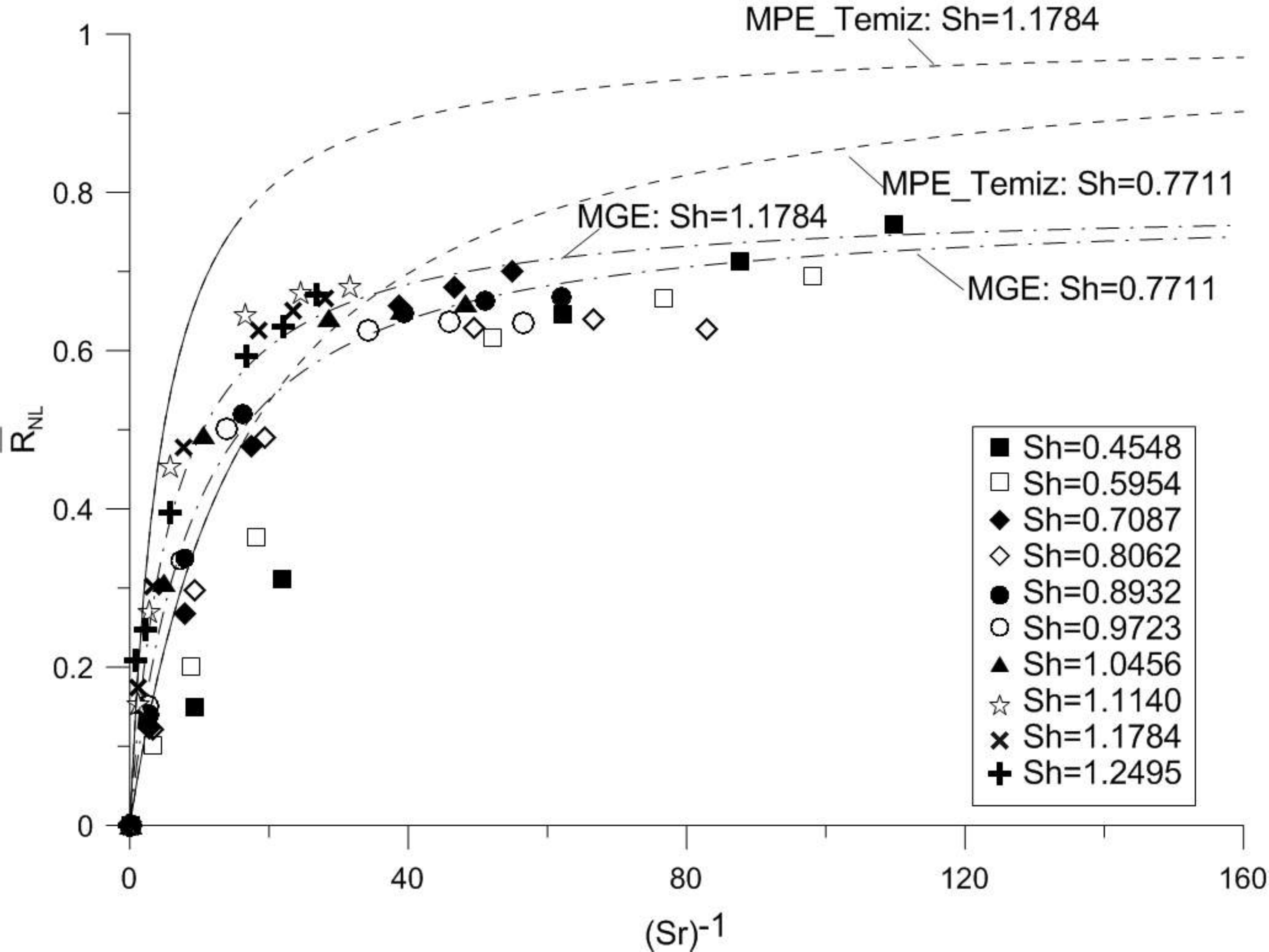}}  
         {\label{fig_NL_rea_MGE1}
          \textbf{b)}\includegraphics[scale=0.3]{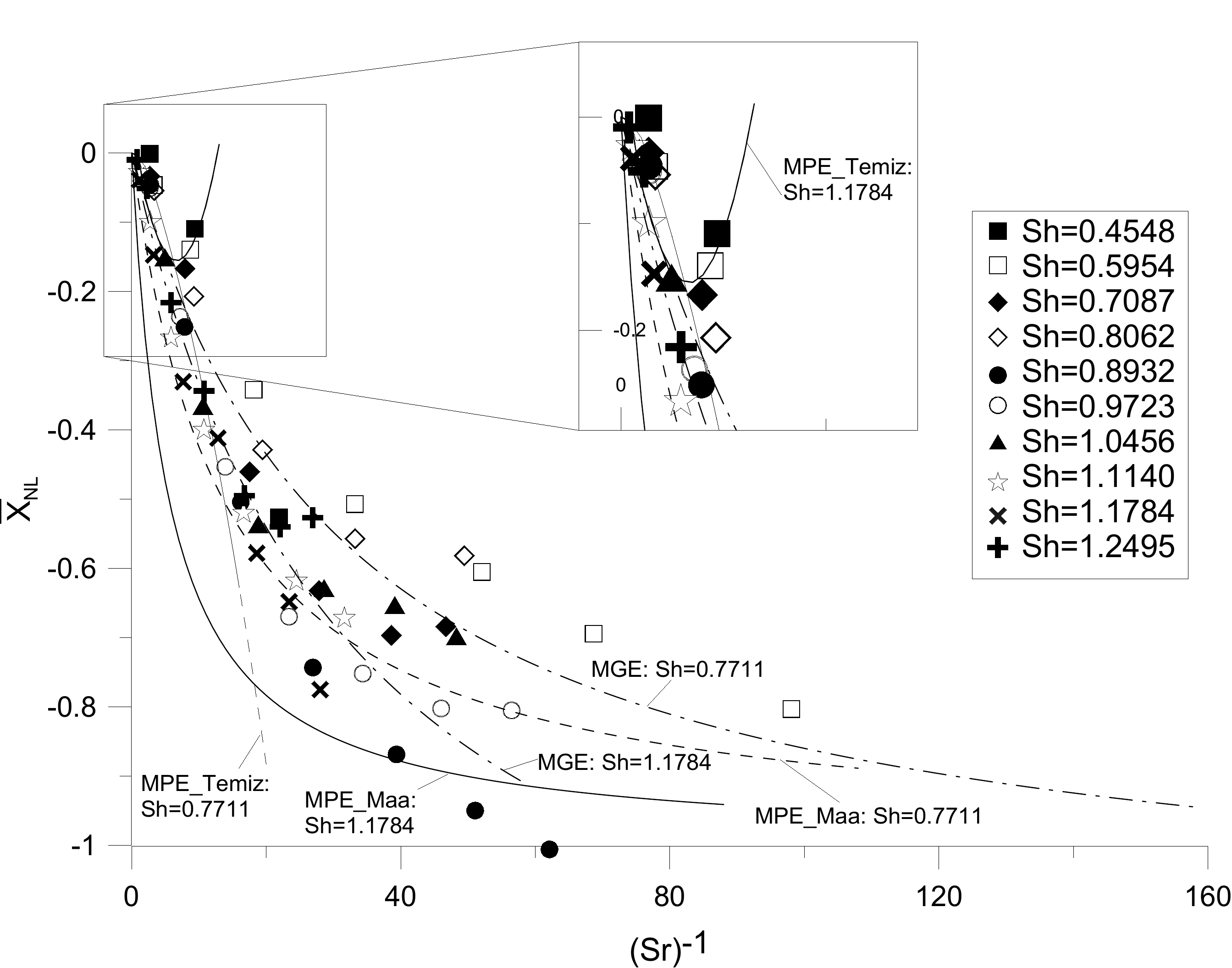}}  
\caption{Normalized non-linear impedance for MGE1 and comparison with MPE: a) Resistive part; b) Reactive part.}  
         \label{fig_NL_imp_MGE1}
\end{flushleft}
\end{figure}    
\subsection{Transfer impedance in quasi- and non- linear regimes}
\begin{figure}
\begin{flushleft}
        {\label{fig_NL_res_MGE3}
          \textbf{a)}\includegraphics[scale=0.25]{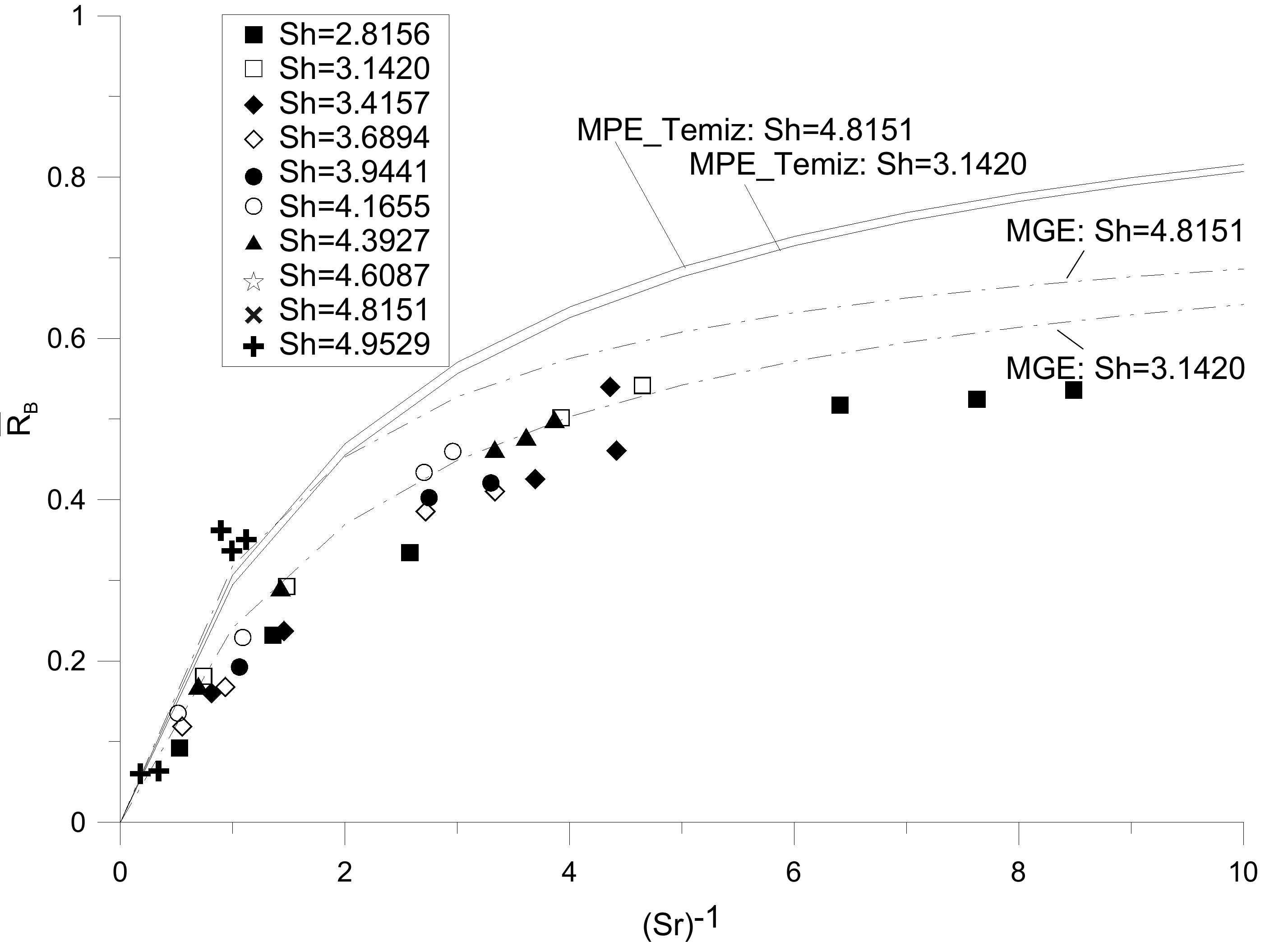}}  
         {\label{fig_NL_rea_MGE3}
          \textbf{b)}\includegraphics[scale=0.25]{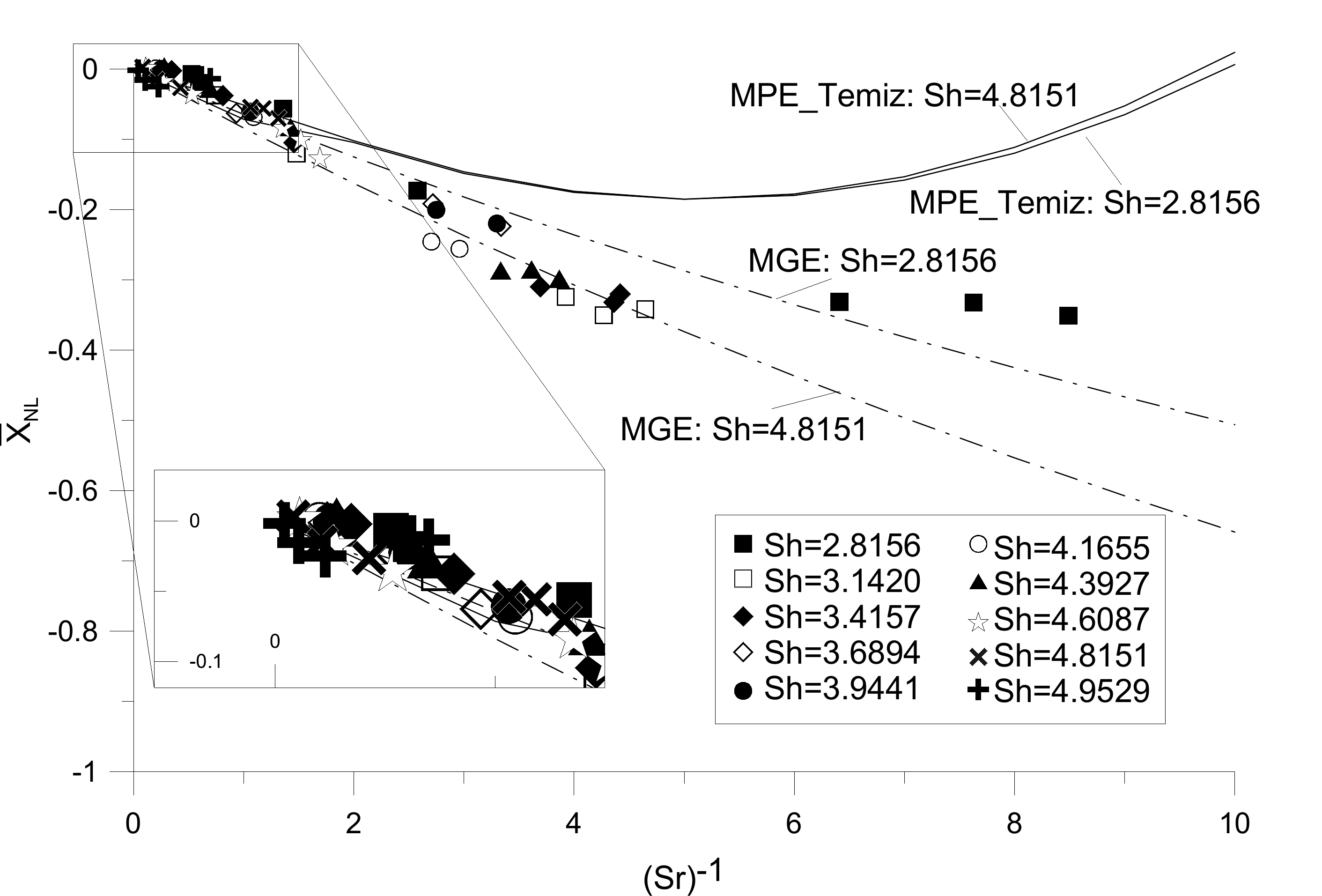}}  
\caption{Normalized non-linear impedance for MGE3 and comparison with MPE: a) Resistive part; b) Reactive part.}  
         \label{fig_NL_imp_MGE3}
\end{flushleft}
\end{figure}

By using the normalizations in Eq.s \ref{Temiz_NL_res} and \ref{Temiz_NL_rea} the results in Fig. \ref{fig_NL_imp_MGE1} are extracted from the data of MGE1. Due to the high level of sound excitation and the small value $d$, the Strouhal number within the apertures is $\ll 1$, resulting in a full non-linear regime.
In order to have a clear picture of the quasi-linear regime, also the data extracted for the MGE3 - provided with larger apertures - are shown in Fig.\ref{fig_NL_imp_MGE3}. 

The following expressions for MGEs have been found by curve fitting the values in Fig.s \ref{fig_NL_imp_MGE1} and \ref{fig_NL_imp_MGE3} :
\begin{equation}\label{MGE_NL_res}
\overline{R}_{NL}=\frac{2.22 \textit{Sr}^{-1}}{2.82 \textit{Sr}^{-1} \textit{Sh}+20}
\end{equation}

\begin{equation}\label{MGE_NL_rea}
\overline{X}_{NL}=\frac{-1.25 \textit{Sh}^{0.5} \textit{Sr}^{-1}+1.283}{\textit{Sr}^{-1}+32.3}-0.0397
\end{equation}\\
By using the Eq.s \ref{Temiz_NL_res} and \ref{Temiz_NL_rea}, it is possible to extract the resistive $R_{NL}$ and reactive part $X_{NL}$ of the non-linear transfer impedance.
In Fig.s \ref{fig_NL_imp_MGE1} and \ref{fig_NL_imp_MGE3} the normalized non-linear transfer impedance of MPEs provided by Temiz in \cite{Temiz_Nonlinear_acou_transfer} is also shown. 
The forumlation presented by Temiz is limited to the \textit{Sr} numbers up to 20 thus, for higher values, the corresponding curves are dashed.  
A noticeable discrepancy appears between the MPEs and the MGEs, mainly in the resistive part. This means that the normalization performed (Eq.s \ref{Temiz_NL_res} and \ref{Temiz_NL_rea}) allows generalizing the transfer impedance for elements of the same typology provided with different values of $d$, $\sigma$ and $C_{vc}$. However, the results obtained for MPEs cannot be used for MGEs, since they represent two different typologies of absorbers.
The values of the normalized resistive part of the non-linear transfer impedance exhibited by MGEs are always lower than the ones exhibited by MPEs. As a consequence, for MGE and MPE provided with the same porosities (which means same particle velocities within the apertures), at increasing of the sound excitation levels the variations in the resistive term is always smaller in case of MGEs than in case of MPEs. It implies that the performance of MGE is more robust than in case of MPE.

The reactive part is more difficult to predict accurately for both MPEs and MGEs. In case of MPEs this part seems to be still higher than in case of MGEs. However, when the resistive part $X_{NL}$ is extracted from the normalized form $\overline{X}_{NL}$, the difference reduces. 
Also in this case, the expression provided by Temiz in \cite{Temiz_Nonlinear_acou_transfer} is reliable up to the transitional regime. In non-linear regime the behaviour predicted by Maa, and expressed by, 
\begin{equation}\label{Maa_NL_rea}
X_{NL-M}=\frac{\chi_{Ch-Ext}}{1+|\frac{u}{\sigma c}|} 
\end{equation}
is acceptable also in case of MGEs (see Fig.\ref{fig_NL_imp_MGE1}b).

\begin{figure}
\begin{flushleft}
        {\label{fig_MGE_vs_MPE_NonLinear_impedance}
          \textbf{a)}\includegraphics[scale=0.3]{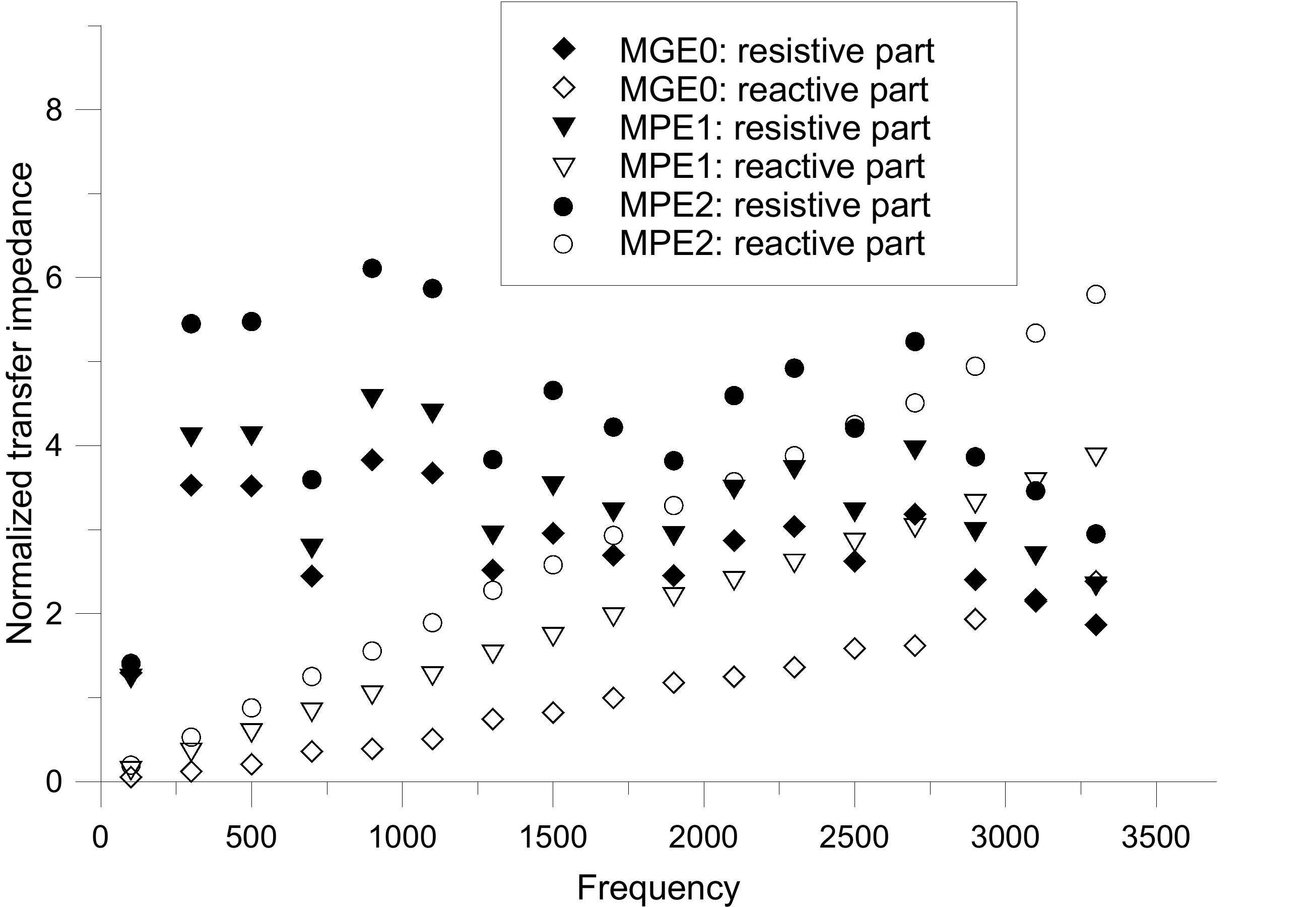}}  
         {\label{fig_MGE_vs_MPE_NonLinear_alpha}
          \textbf{b)}\includegraphics[scale=0.3]{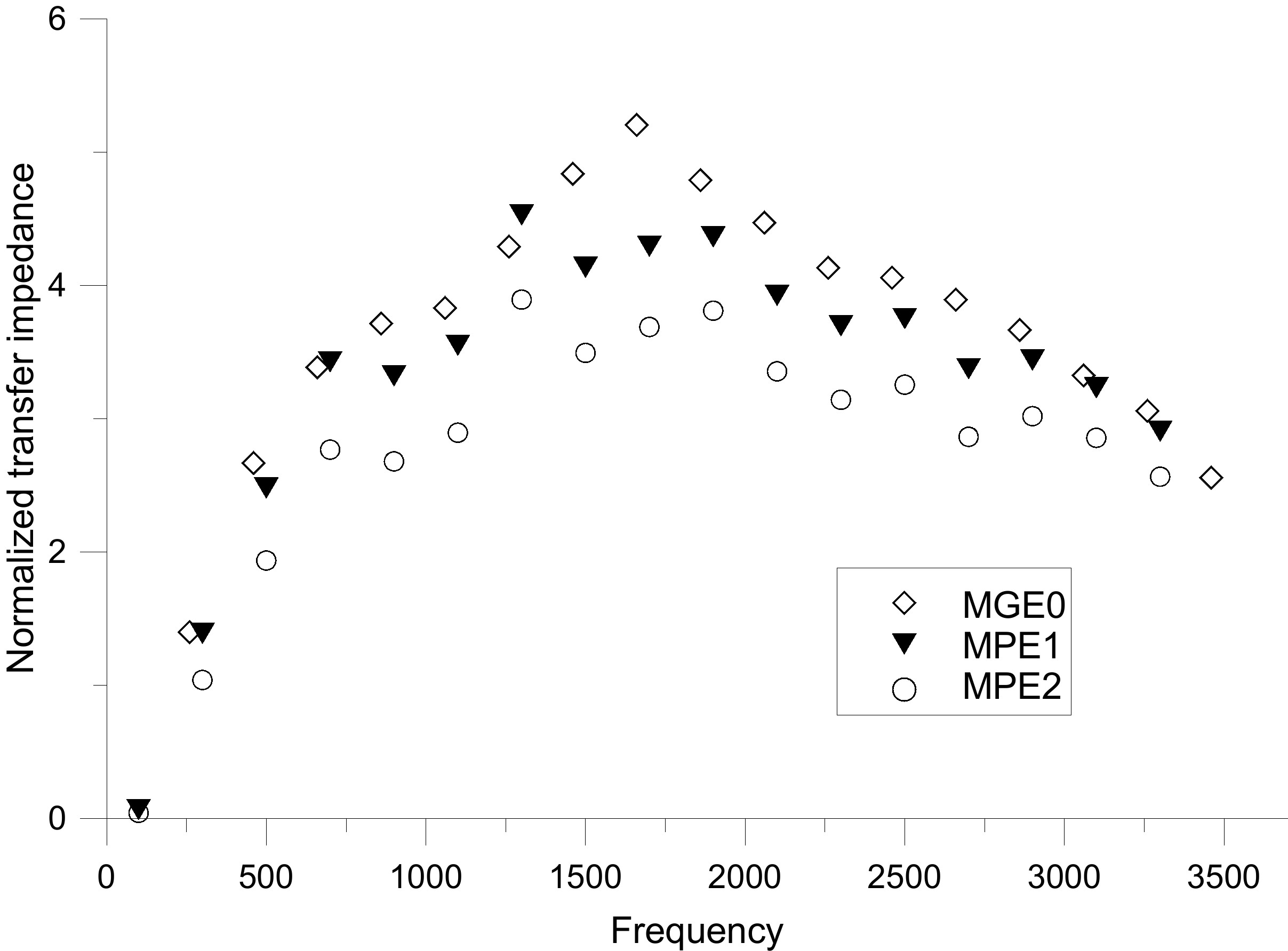}}  
\caption{Non-linear behaviour of MGEs and MPEs: a) Transfer impedance; b) Absorption coefficients.}  
         \label{fig_MGE_vs_MPE_Non_linear}
\end{flushleft}
\end{figure} 
\subsection{Absorption coefficient in non-linear regime}
The resistive and reactive components, as well as the absorption coefficients plotted in Fig. \ref{fig_MGE_vs_MPE_Non_linear}, refer to the same MGE and MPE samples plotted in Fig.\ref{fig_MGE_vs_MPE_linear} but tested with high levels of sound excitation. The performance of MGE0 and MPE1, in this case, show an advantage in the absorption coefficient of the latter over the last one of $\approx 6\%$. 
Moreover, the absorption coefficient of the MGE0 has an advantage of $\approx 21 \%$ over the traditional MPE2. Since the samples MPE1 and MGE0 are provided with similar porosities, the particle velocities averaged in the aperture areas, as well as the \textit{Sr} in MGE0 and in MPE1, are similar too. 
However, by comparing Fig.\ref{fig_MGE_vs_MPE_linear_impedance}a with Fig.\ref{fig_MGE_vs_MPE_Non_linear}a it is possible to notice that high excitation levels result in  greater variation of the resistance in case of MPE1 than in case of MGE0. As a consequence, the decay of absorption peak for MGE0 is smaller than for MPE1.
\subsection*{Random noise excitation}
\begin{figure}
\begin{flushleft}
        {\label{fig_WN_NL_Res_MGE}
          \textbf{a)}\includegraphics[scale=0.3]{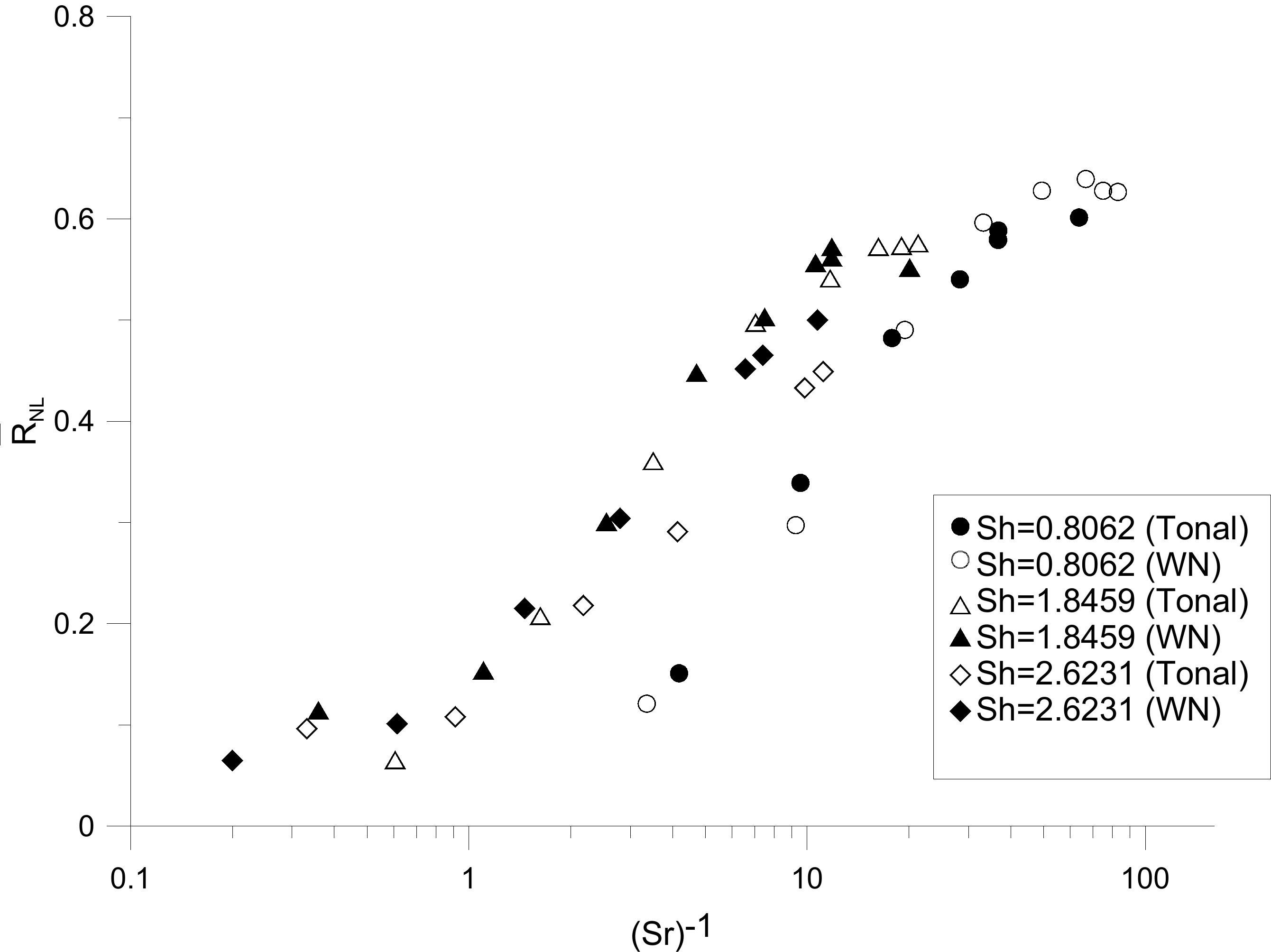}}  
         {\label{fig_WN_NL_Rea_MGE}
          \textbf{b)}\includegraphics[scale=0.3]{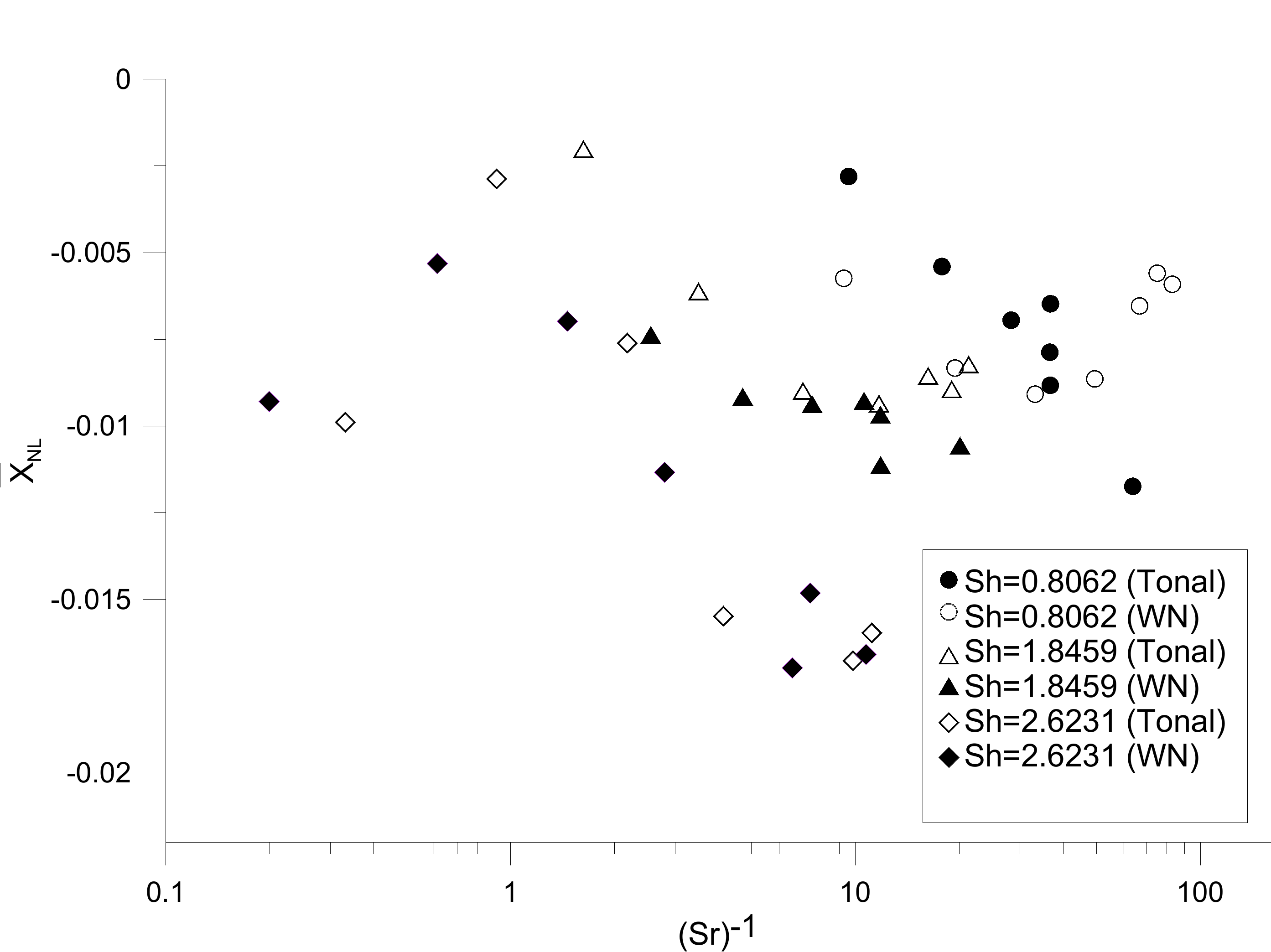}}  
\caption{Normalization of data extracted with white noise excitation: a) Non-linear resistance; b) Non-linear reactance.}  
         \label{fig_WN_NL_Imp_MGE}
\end{flushleft}
\end{figure}

As a conclusive note, the MGE samples have been tested also in presence of random noise, in the frequency range $ \left[ 100, 3700 \right] Hz$ and at different levels of sound excitation. In order to model the acoustic behaviour of MGE in these conditions, the following fictitious velocity can be defined
\begin{equation}\label{WN_conversion}
u^*=2.3 \sqrt{\sum\limits_{i=F_s}^{F_f} \tilde{u}_i^2}
\end{equation}
where the sum is calculated from the first ($F_s$) to the final ($F_f$) frequency of the spectrum and $\tilde{u}$ is the \textit{rms} of the particle velocities within the apertures.
By using $u^*$ instead of $u$ in the definition of \textit{Sr} (Eq.\ref{Shear_number} ) and in the normalization Eq.\ref{Temiz_NL_res}, the resistive part of the non-linear transfer impedance, measured in presence of random noise excitation, can be still expressed by using the Eq.\ref{MGE_NL_res}. In fact, the normalized resistance obtained by using single tones and particle velocity $u$ and the normalized resistance obtained by using random noise and the velocity $u^*$, collapse well in Fig. \ref{fig_WN_NL_Imp_MGE}a.

However, the normalization Eq.\ref{Temiz_NL_rea} does not include explicit dependence on the particle velocity. Thus, in order to use Eq. \ref{MGE_NL_rea}, $u^*$ must be used instead of $u$ to compute \textit{Sr} and the following normalization must be used to express the reactance:

\begin{equation}\label{NL_rea_NormD}
\Delta \chi_{NL-B}=\frac{\overline{X}_{NL} \sigma}  { \frac{\chi_{Ch-Ext}|\frac{u}{\sigma c}|}{1-|\frac{u}{\sigma c}|}  }
\end{equation}

Fig. \ref{fig_WN_NL_Imp_MGE}b show that also the data of the reactance, obtained by using sine excitation and normalized according to Eq. \ref{NL_rea_NormD}, and the ones obtained by using the white noise, $u^*$ instead of $u$ and the same normalization, collapse well.

\section{Conclusions}
The ideal requirement in micro-perforated element of having infinitely small and short apertures, while still preserving adequate porosity, sufficient mechanical properties and cost effective manufacturing process, is targeted by micro-grooved elements. By means of two mating layers obtained with basic production technology, it is possible to generate a number of micro-channels with depth in the order of $100 \mu m$ and length of $200 \mu m$, thus providing high acoustic performance. The thickness of the layer is responsible for mechanical properties and has a minor effect on the acoustic performance. 

In the linear regime, the micro-channels act as slit shaped apertures with reduced resistive end-resistance. The absorption coefficient of a micro-grooved element provided with $\approx 3$ micro-channels per square centimetre with depth $95 \mu m$ and length $200 \mu m$, is $\approx 10\%$ higher than the absorption coefficient of a micro-perforated element with $\approx 22$ holes per square centimetre with $300 \mu m$ diameter of perforation and thickness of $500 \mu m$. The reactive end-correction of MGE has been described by extending to sub-millimetre apertures the formulation provided in literature for over-millimetre apertures.

Compared to micro-perforated element, the quasi- and non- linear behaviour of micro-grooved element is characterized by reduced variation of the resistive transfer impedance at increasing the sound excitation levels. Thus, more robust acoustic performance is exhibited. A semi-empirical approach, obtained by curve fitting the normalized experimental data, has been used to describe the non-linear transfer impedance of micro-grooved element as a function of Shear number, Strouhal number and discharge coefficient. By properly defining a fictitious particle velocity within the apertures, the same formulation - originally developed for pure tone excitation - has been extended to the case of random excitation noise.    

\section{Acknowledgement}
The author would like to acknowledge Tallinn University of Technology Development Found, object SS346, for the financial support.

\bibliographystyle{elsarticle-num}
\section*{References}

\bibliography{Manuscript}

\begin{thebibliography}{10}
\expandafter\ifx\csname url\endcsname\relax
  \def\url#1{\texttt{#1}}\fi
\expandafter\ifx\csname urlprefix\endcsname\relax\def\urlprefix{URL }\fi
\expandafter\ifx\csname href\endcsname\relax
  \def\href#1#2{#2} \def\path#1{#1}\fi

\bibitem{Liu_Architectural_acoustics}
K.~Liu, J.~Tian, X.~Li, F.~Jiao, An ideal absorbing materials in architectural
  acoustics: Micro‐perforated panel absorbers, Journal of Acoustic Society of
  America 119.

\bibitem{Kabral_Compact_nonfibrous_silencer}
R.~Kabral, L.~Du, M.~Knutsson, Optimization of compact non-fibrous silencer for
  the control of compressor noise, SAE Technical Paper 2016-01-1818.

\bibitem{Bielak_Turbofan_Liner}
G.~W. Bielak, J.~W. Premo, A.~S. Hersh, Advanced turbofan duct liner concepts,
  Tech. rep., NASA, prepared for Langley Research Center Hampton, Virginia
  (February 1999).

\bibitem{Allam_A_new_type_of_muff}
S.~Allam, M.~\r{A}bom, A new type of muffler based on microperforated tubes,
  Journal of Vibration and Acoustics 133 (2011) 1--8.

\bibitem{MGE_SETC_Application_to_small_eng_silencer}
F.~Auriemma, H.~R\"ammal, J.~Lavrentjev, Application of novel micro-grooved
  elements to small engine silencer, SAE Technical paper JSAE 20139001 (2013)
  1--10, SAE Japan.

\bibitem{Maa5_Basic_theory_of_acoustics}
D.~Y. Maa, Basic theory of acoustics, Beijing, 2003, chapter 10 (Chinese).

\bibitem{Maa2_Theory_and_design_of_microperf}
D.~Y. Maa, Theory and design of micro perforated - panel sound absorbing
  construction, Sci. Sin. 18 (1975) 55--71.

\bibitem{Maa3_Microperf_at_high_sound}
D.~Y. Maa, Microperforated panel at high sound intensity, Proceedings of
  Internoise 94 (1994) 1--8.

\bibitem{Allard_book}
J.~F. Allard, N.~Atalla, Propagation of sound in porous media – Modelling
  sound absorbing materials, 2nd Edition, John Wiley \& Sons, 2009.

\bibitem{Liu_Enhancing_by_partitioning_cavity}
J.~Liu, D.~W. Herrin, Enhancing micro-perforated panel attenuation by
  partitioning the adjoining cavity, Applied Acoustics 71 (2010) 120--127.

\bibitem{Sagakami_Enhancing_by_multiple_leaf}
K.~Sagakami, M.~Yairi, M.~Morimoto, Multiple-leaf sound absorbers with
  microperforated panels: an overview, Acoustics Australia 38 (2010) 76--81.

\bibitem{Park_microperf_backed_by_Helmholtz}
S.~H. Park, Acoustic properties of micro-perforated panel absorbers backed by
  helmholtz resonators for the improvement of low-frequency sound absorption,
  Journal of Sound and Vibration 332 (2013) 4895--4911.

\bibitem{Quian_Ultra_micro_perforations}
Y.~J. Quian, D.~Y. Kong, S.~M. Liu, S.~M. Sun, Z.~Zhao, Investigation on
  micro-perforated panel absorber with ultra-micro perforations, Journal of
  sound and Vibration 74 (2013) 931--935.

\bibitem{Randeberg_Perforated_absorbers}
R.~T. Randeberg, Perforated panel absorbers with viscous energy dissipation
  enhanced design, Ph.D. thesis, Trondheim, Norway (June 2000).

\bibitem{Knipstein_Acoustimet}
D.~M. Knipstein, Sound absorbing element and procedure for manufacture of this
  element and use of this element, Tech. Rep. patent number EP0876539B1,
  European patent Office.

\bibitem{Pfretzchner_Insertion_units}
J.~Pfretzschner, A.~Fernandez, P.~Cobo, M.~Cuesta, A.~Fernandez,
  Microperforated insertion units: An alternative strategy to design
  microperforated panels, Applied Acoustics 67~(1) (2006) 62--73.

\bibitem{AuriemmaMGE1_A_novel_solution}
F.~Auriemma, H.~R\"ammal, J.~Lavrentjev, A novel solution for noise control,
  SAE International Journal of Materials and Manufacturing 3 (2013) 599--610,
  doi:10.4271/2013-01-1941.

\bibitem{Melling_Acoust_impedance_at_high_medium_pressure}
T.~Melling, The acoustic impedance of perforates at medium and high sound
  pressure levels, Journal of Sound and Vibration 29 (1973) 1--65.

\bibitem{Cummings_High_amplitude_transmission}
A.~Cummings, W.~Eversman, High amplitude acoustic transmission through duct
  terminations: theory, Journal of Sound and vibration 91 (1983) 503--518.

\bibitem{Elnady_Semi_Empirical_liner_impedance_modeling}
T.~Elnady, H.~Bod\'{e}n, On semi-empirical liner impedance modeling with
  grazing flow, 9th AIAA/CEAS Aeroacoustics Conference and Exhibit, 12-14 May
  2003, Hilton Head, South California.

\bibitem{Temiz_Nonlinear_acou_transfer}
M.~A. Temiz, J.~Tournadre, I.~L. Arteaga, A.~Hirschberg, Non-linear acoustic
  transfer impedance of micro-perforated plates with circular orifices, Journal
  of Sound and Vibration 366 (2016) 418--428.

\bibitem{Park_microperf_launcher_fairings}
S.~H. Park, A design method of micro-perforated panel absorber a thigh sound
  pressure environment in launcher fairings, Journal of Sound and Vibration 332
  (2013) 521--535.

\bibitem{Chandrasekharan_Acoustic_impedance_of_MEMS_based_microperf}
V.~Chandrasekharan, M.~Sheplak, L.~Cattafesta, Experimental study of acoustic
  impedance of mems-based micro-perforated liners, 12th AIAA/CEAS Aeroacoustics
  Conference, 08-10 May 2006, Cambridge, Massachussets.

\bibitem{Majak_genetic_algoritm}
J.~Majak, M.~Pohlak, M.~Eerme, Design of car frontal protection system using
  neural networks and genetic algorithm, Mechanika 4 (2012) 453--460.

\bibitem{Pierce_book}
A.~D. Pierce, Acoustics - An introduction to its physical principles and
  applications, 2nd Edition, Acoustical Society of America through the American
  Institute of Physics, 1994.

\bibitem{Lavrentjev_A_measurement_method}
J.~Lavrentjev, M.~\r{A}bom, H.~Bod\'{e}n, A measurement method for determining
  the source data of acoustic two-port sources, Journal of Sound and Vibrations
  183~(3) (1995) 517--531.

\bibitem{Chung_Transfer_funct}
J.~Y. Chung, D.~A. Blaser, Transfer function method of measuring in-duct
  acoustic properties. i. theory, Journal of the Acoustical Society of America
  68 (1980) 907.

\bibitem{Munjal_book}
M.~L. Munjal, Acoustics of ducts and mufflers, 2nd Edition, John Wiley \& Sons,
  2014.

\bibitem{Abom_Error_analysis}
M.~\r{A}bom, H.~Bod\'{e}n, Error analysis of two-microphone measurements in
  ducts with flow, Journal of the Acoustical Society of America 83 (1988)
  2429--2438.

\bibitem{Crandall_Theory}
I.~Crandall, Theory of vibrating systems and sound, D. Van Nostrand \& Co.
  Inc., New York, 1927.

\bibitem{Temiz_End_corrections}
M.~A. Temiz, I.~L. Arteaga, G.~Efraimsson, M.~\r{A}bom, A.~Hirschberg, The
  influence of edge geometry on end-correction coefficients in micro perforated
  plates, Journal of the Acoustical Society of America 138.

\bibitem{Ingard_Acoustic_resonators}
U.~Ingard, On the theory and design of acoustic resonators, Journal of the
  Acoustical Society of America 25 (1953) 1037--1062.

\bibitem{Disselhorst_Almost_linear_regime}
J.~H.~M. Disselhorst, L.~van Wijngaarden, Flow in the exit of open pipes during
  acoustic resonance, Journal of Fluids and Structures 99.

\bibitem{Hofmans_Aeroacoustic_resp_of_slit_quasi_steady}
G.~C.~J. Hofmans, R.~J.~J. Boot, P.~P. J.~M. Durrieu, Y.~Aur\'{e}gan,
  A.~Hirshberg, Aeroacoustic response of a slit-shaped diaphragm in a pipe at
  low helmholts number, 1: quasi-steady results, Journal of sound and Vibration
  244 (2001) 35--56.

\end{thebibliography}


\end{document}